\documentclass[reprint,amsmath,amssymb,graphicx]{revtex4-1}

\usepackage{graphicx}

\begin{document}

\preprint{APS/123-QED}

\title{Simulation of fluid-solid coexistence in finite volumes: A method to study the properties of wall-attached crystalline nuclei}

\author{Debabrata Deb, Alexander Winkler, Peter Virnau, and Kurt Binder}
\affiliation{Institut f\"ur Phsyik, Johannes Gutenberg-Universit\"at Mainz\\
 Staudingerweg 7, D-55099 Mainz, Germany} %



\date{\today}

\begin{abstract}
The Asakura-Oosawa model for colloid-polymer mixtures is studied by Monte Carlo simulations at densities
inside the two-phase coexistence region of fluid and solid. Choosing a geometry where the system is confined
between two flat walls, and a wall-colloid potential that leads to incomplete wetting of the crystal at the wall,
conditions can be created where a single nanoscopic wall-attached crystalline cluster coexists with fluid in the
remainder of the simulation box. Following related ideas that have been useful to study heterogeneous nucleation of
liquid droplets at the vapor-liquid coexistence, we estimate the contact angles from observations of the crystalline clusters
in thermal equilibrium. We find fair agreement with a prediction based on Young's equation, using estimates of interface
and wall tension from the study of flat surfaces. It is shown that the pressure versus density curve of the finite
system exhibits a loop, but the pressure maximum signifies the ``droplet evaporation-condensation'' transition and thus
has nothing in common with a van der Waals-like loop. Preparing systems where the packing fraction is deep inside the 
two-phase coexistence region, the system spontaneously forms a ``slab state'', with two wall-attached crystalline domains 
separated by (flat) interfaces from liquid in full equilibrium with the crystal in between; analysis of such 
states allows a precise estimation of the bulk equilibrium properties at phase coexistence.
\end{abstract}

\maketitle


\section{INTRODUCTION AND OVERVIEW}
Nucleation of crystals from fluid phases is an important problem \cite{1,2,3,4,5} with important
applications, such as formation of ice crystals in the atmosphere, solidification of molten
silicates in processes deep underneath the earth crust, and last but not least crystallization processes
of various materials are ubiquitous in many technical processes. However, nevertheless crystal
nucleation is rather poorly understood on a quantitative level: one mostly relies on the concept of classical
nucleation theory \cite{2,6,7,8,9,10}, but since almost always the ``critical nucleus'' that triggers the
phase transition contains only a few tens to at most a few thousand particles, considerations based on
macroscopic concepts (balancing bulk and surface free energies, using the interfacial tension of macroscopic
flat interfaces, etc. \cite{1,2,3,4,5}) are doubtful. Moreover, in most cases of interest nucleation is not
homogeneous (i.e., triggered by spontaneous thermal fluctuations) but rather heterogeneous \cite{9,11,12,13,14,15}
(i.e., triggered by defects, e.g. nucleation of ``droplets'' attached to a flat solid wall facilitating formation of
crystal planes parallel to the wall stacking on top of each other). Again, macroscopic concepts (involving
the ``contact angle'' \cite{16,17,18} at the wall, and possibly a free energy excess due to the three-phase
contact line where the crystal and the fluid meet at the wall, the ``line tension'' \cite{16,19,20,21} are used
\cite{22}, but their reliability is uncertain.

Recently progress has been achieved by studying the nucleation of colloidal crystals \cite{23,24,25,26,27}.
Colloidal particles are in the $\mu m$ range, and hence the structure of fluid-crystal interfaces can be
studied with single-particle resolution \cite{28}, and since dynamics of such systems are very slow, the time
evolution of interfacial phenomena can be followed in real time \cite{29,30}. A further bonus is that effective
interactions between colloidal particles are tunable to a large extent \cite{31,32,33}. A good example of this
point are colloid-polymer mixtures \cite{34,35,36}: varying the polymer concentration in the colloidal dispersion
one can vary the depletion attraction between the particles \cite{35}.

\begin{figure} [ht]
\includegraphics[scale=0.290]{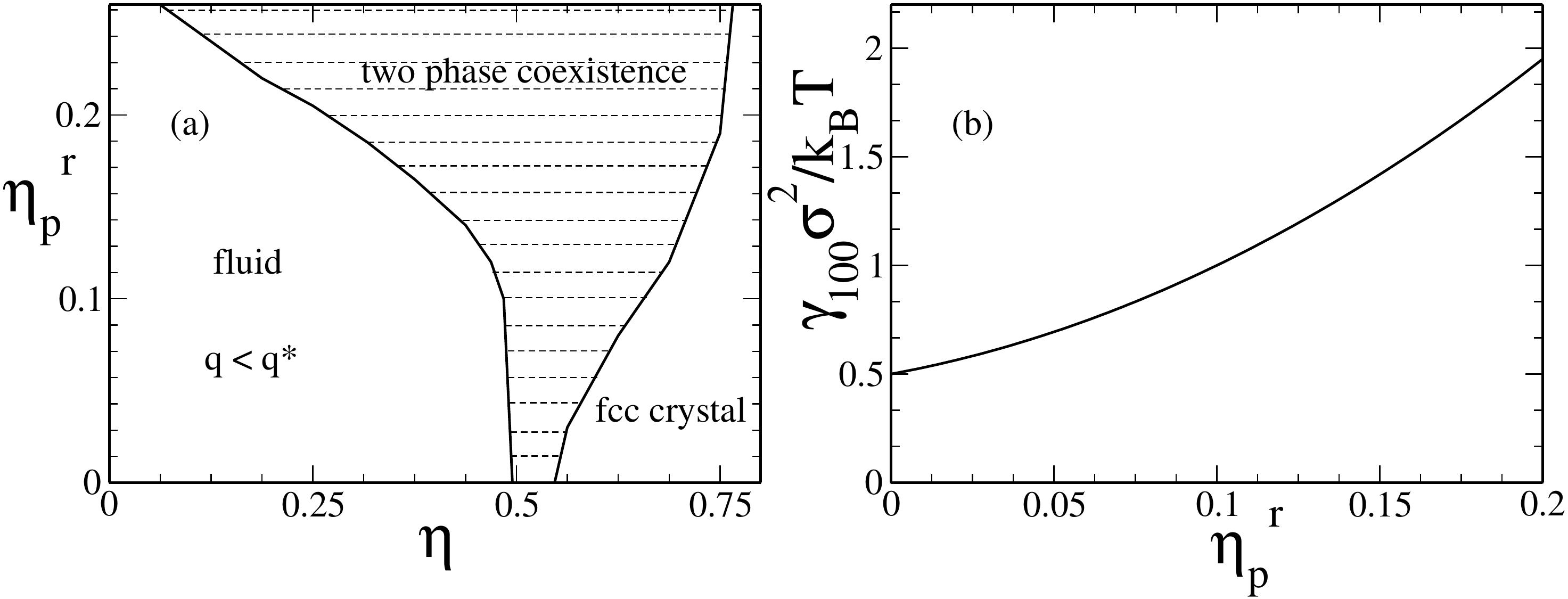}
\caption{\label{fig1} Schematic phase diagram of the Asakura-Oosawa model for a colloid polymer mixture in the plane
of variables polymer reservoir packing fraction $\eta_p^r$ and colloid packing fraction $\eta$ (left part), and
variation of the fluid-solid interface tension $\gamma_{100} \sigma^2/k_ BT$ vs. $\eta_p^r$ (right part). Here $\sigma$
is the colloid diameter, $\sigma_p$ the polymer diameter, $q = \sigma_p/\sigma$ and $q*=0.154$ the critical value of
$q$, above which three-body interactions appear in the effective interaction between colloids mediated by the
polymers. The region of two-phase coexistence, in between the volume fraction $n_f$ (where freezing begins) and
$\eta_m$ (where melting begins) has been shaded. Note that for fluid-solid interfaces the interface tension depends
on the interface orientation, $\gamma_{100}$ refers to an interface that is perpendicular to the $x$-axis.}
\end{figure}

This effect has first been described in term of a simple model, the Asakura-Oosawa model \cite{37,38,39}, and subsequently
it has been shown that this model does account qualitatively for the experimental observations very well
\cite{35,36}. Thus, varying the polymer concentration the width of the two-phase coexistence region
and the magnitude of the interfacial tension between coexisting fluid and solid phases can be controlled
\cite{40}, as sketched in Fig.~\ref{fig1}. If the colloid-polymer mixture is confined between two
(equivalent) walls, and the walls are prepared such that there is incomplete wetting \cite{16,17,18} of the solid at the walls,
it is likely that by variation of the polymer concentration in the dispersion one can change the contact angle.
Such a control of wetting properties is very difficult to achieve in small molecule systems.

\begin{figure} [ht]
\includegraphics[scale=0.30]{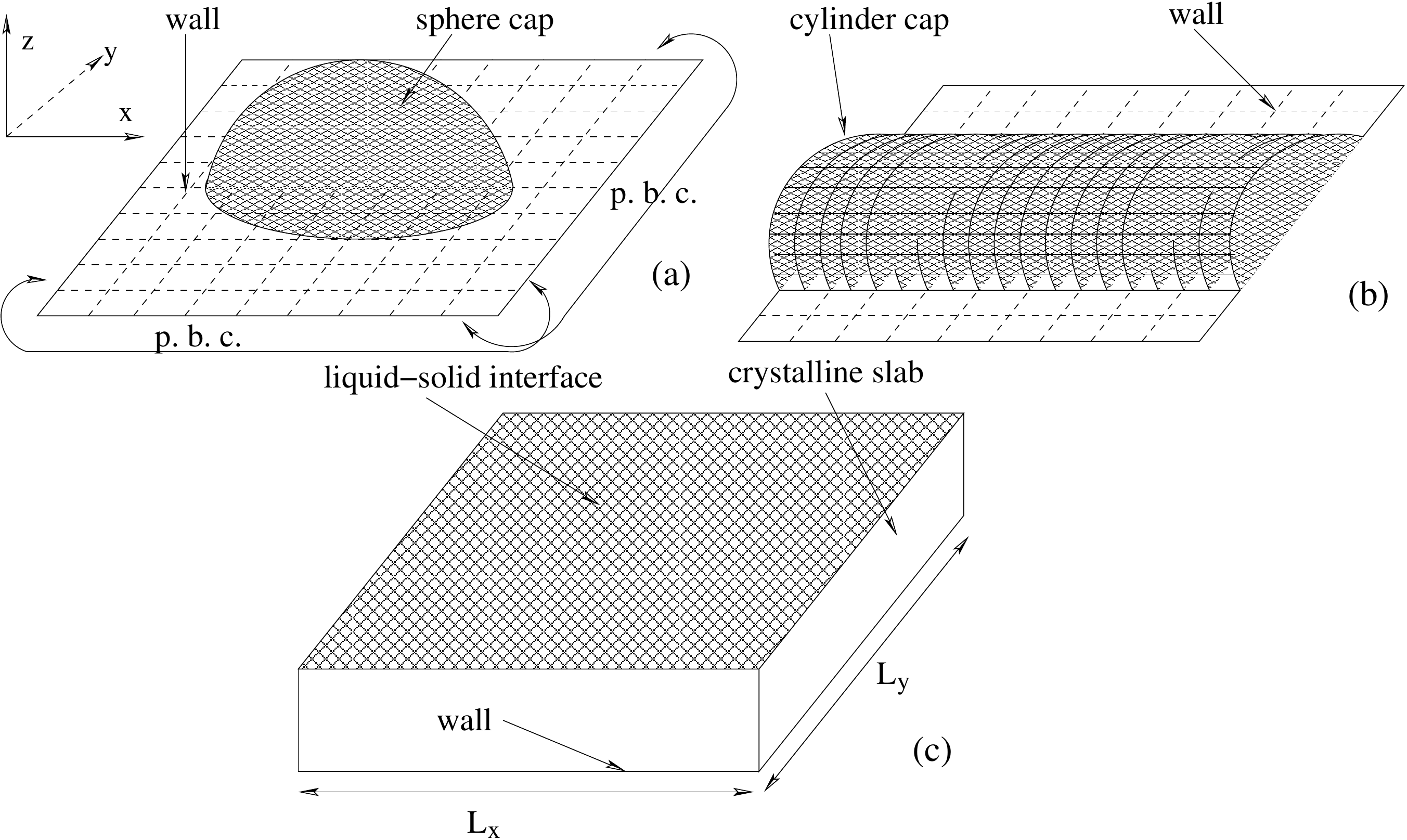} 
\caption{\label{fig2} Schematic description of two-phase configurations in a thin film geometry of volume $L_x \times
L_y \times D$ geometry, with periodic boundary conditions in $x$- and $y$-directions, and two walls at $z=0$ and
$z=D$. It is assumed that the wall at $z=0$ exhibits incomplete wetting by the crystal, while the wall at $z=D$
is not shown. All three cases shown assume a colloid packing fraction $\eta$ in the two phase coexistence
region chosen such that the solid phase is the minority phase. For $\eta_t < \eta < \eta_{t'}$ the solid domain
is a sessile sphere-cap-shaped crystalline cluster attached to the wall (a) while for $\eta_{t'} < \eta < \eta_{t''} $ it is a
cylinder cap (b), and for $\eta > \eta_{t''}$ it is a crystalline slab (c). Note that cases (a,b) ignore the
crystalline structure of the solid, which rather is treated in a continuum description, ignoring both the
anisotropy and discreteness of the lattice structure} \end{figure}

In the present paper we hence want to contribute to the theoretical understanding of wall-attached crystalline
clusters in colloid-polymer mixtures by computer simulation methods. As is well known, for macroscopic systems
the ``critical droplets'' occurring in nucleation processes are hard to observe, since one has to focus on
transient rare events when the system traverses a saddle point in the free energy landscape \cite{41,42,43},
when a droplet grows from subcritical to supercritical size. Here, computer simulations possess an advantage because
in a system of conserved density in a finite-sized simulation box the coexistence of the critical droplet with surrounding
``parent'' phase is a situation of stable equilibrium \cite{44}, unlike the situation in the thermodynamic limit
where the ``parent'' phase is metastable, and the droplet on top of the saddle is in unstable equilibrium \cite{43}.
In previous work, it has been shown that critical droplets (or bubbles, respectively) associated with the liquid-vapor
transition \cite{45,46} or systems undergoing an unmixing transition in symmetric binary fluids
\cite{46,47} or Ising models \cite{14,15,48} can be studied in this way. When the packing fraction ($\eta$) of the finite system is chosen
such that it falls inside of the two-phase coexistence region of the infinite system, i.e.~$\eta_f < \eta < \eta_m$
(Fig.~\ref{fig1}), we may encounter phase coexistence inside the simulation box. For a fluid-solid transition in
thin film geometry with walls where incomplete wetting by the crystal occurs, we expect various shapes of the minority
domain, depending on $\eta$ as shown in Fig.~\ref{fig2}. Of course, Fig.~\ref{fig2} is inspired by analogous studies of
vapor-liquid transitions \cite{14,15}, where both coexisting fluid phases are homogeneous and isotropic: only then it does
 clearly make sense to describe the minority domain as sphere caps (Fig.~\ref{fig2}a) or cylinder caps (Fig.~\ref{fig2}b)
that are stabilized by the periodic boundary condition (and oriented in the $x$-direction when $L_x < L_y$ and along
the $y$-direction when $L_x > L_y$, while for $L_x=L_y$ there occurs a degeneracy). For solid-liquid coexistence
in finite volumes, Fig.~\ref{fig2} is approximate because of two reasons: (i) on the nanoscale, when the height
of the sphere cap or cylinder cap is only a few lattice spacings, the discrete lattice structure of the crystal
should be taken into consideration (ii) only when the linear dimensions of the crystalline domain are very much larger
than the lattice spacing and therefore the question what is its ``macroscopic shape'' makes sense. But even then Figs.~\ref{fig2}a), b) only hold
when the fluid-solid interface tension does not depend on the orientation of the interface. Already in the bulk
the shape of a crystal in general therefore is never a sphere, but rather needs to be found from the anisotropic interface
tension via the Wulff construction \cite{49,50,51}. The extension of this construction to wall-attached crystals
has been given by Winterbottom et al. \cite{52,53,54}. In fact, it is a nontrivial question under which conditions
planar facets (rather than curved interfaces) occur \cite{55}. Of course, for nano-crystallites we also expect a finite-size rounding of faceting transitions, related to the finite-size rounding of the interfacial roughening transition
\cite{56}, and hence the analysis of the equilibrium shapes of nano-crystals is very subtle. In Fig.~\ref{fig2}, we have
also assumed that the geometry (and conditions at the wall that occurs at $z=D$) can be chosen such that no interfaces occur
that connect both walls (though in fluid systems the occurrence of ``liquid bridges'' is extremely common, see e.g.~
\cite{57,58,59,60,61}). Thus, the present study can present first exploratory steps only.

In Sec. II, we shall define precisely the model that is simulated and give details on the techniques of ``system
preparation'', simulation methods, and analysis techniques (note that it is a delicate matter to decide which particles
are to be counted as part of the crystalline solid domain or as part of the fluid in each microstate of the simulation
\cite{40}). In Sec. III we describe our numerical results and discuss them in the light of the questions that have been
outlined above, while Sec. IV summarizes our conclusions. In the Appendix A, we consider the coexistence between wall-attached crystalline films and a fluid state in between, separated from the crystalline layers by planar interfaces. This is a ``self-regulating'' system, where the thickness of the fluid phase adjusts itself, so that the lever rule holds. We show that this geometry is useful for a direct estimation of the bulk packing fractions $\eta_f$, $\eta_m$ and the coexistence pressure $p_{co}$. In the appendix B, we compare the metastable crystallites (with 100 faces adjacent to the wall)  to stable ones (where the close-packed 111 faces are adjacent to the wall).

\section{MODEL, SIMULATION AND ANALYSIS TECHNIQUES}

The Asakura-Oosawa model of colloid-polymer mixtures \cite{37,38,39} describes the colloids as hard spheres of
diameter $\sigma$, the polymers are described as soft spheres of diameter $\sigma_p$, and both colloid-colloid and
colloid-polymer overlap is strictly forbidden, while polymer-polymer overlap does not cost any energy. Thus, the
interaction potentials are (``c'' stands for colloids, ``p'' for polymers, $r$ is the distance between the
particles)

\begin{equation} \label{eq1}
U_{cc} (r \leq \sigma) = \infty \quad , \quad U_{cc} ( r > \sigma)=0
\end{equation}

  \begin{equation} \label{eq2}
  U_{cp} ( r \leq (\sigma + \sigma _p)/2) = \infty \, ,  \quad U_{cp} (r > (\sigma + \sigma_p)/2)=
  0\, ,
  \end{equation}

  and

  \begin{equation} \label{eq3}
  U_{pp}(r) =0 \quad .
  \end{equation}

  The packing fraction of polymers $(\eta_p)$ and colloids $(\eta)$ are then defined in terms of
  the corresponding densities $\rho_p=N_p/V$ and $\rho_c=N_c/V$ of these particles ($V$ is the total
  volume, $N_p$, $N_c$ are the particle numbers of polymers and colloids, respectively)

  \begin{equation} \label{eq4}
  \eta_p=(\pi \sigma_p^3/6) \rho_p \quad , \quad \eta=(\pi \sigma^3 / 6) \rho_c \quad .
  \end{equation}

  It is convenient to use the chemical potential $\mu_p$ of the polymers as an external control
  variable, or, equivalently, the fugacity $z_p=\exp (\mu_p/k_BT)$. The ``polymer reservoir
  packing fraction'' $\eta^r_p$ then is defined as \cite{32,33,34,35,36,37,38,39}

  \begin{equation} \label{eq5}
  \eta_p^r \equiv (\pi \sigma^3_p/6) z_p \quad .
  \end{equation}

  Being interested in static equilibrium properties of the model, one can proceed by first integrating out all
  the coordinates of the polymers, keeping only the coordinates of the colloids in the system as variables.
  For $q=\sigma_p/\sigma  < q^*=0.154$, this can be done explicitly and shown to yield an effective colloid-colloid attraction
  \cite{62,63}

  \begin{eqnarray} \label{eq6}
  &&U_{cc} (r) / k_BT = - (q^{-1} + 1)^3 \eta^r_p \Big[1-\frac{3r}{2 \sigma (1 +q) } \nonumber\\
  &&+\frac{r^3}{2 \sigma^3 (1 + q)^3}\Big], \quad \sigma < r < \sigma + \sigma_p\,
  \end{eqnarray}

  \begin{equation} \label{eq7}
  U_{cc} (r \geq \sigma + \sigma_p)=0 \quad.
  \end{equation}

  Of course, for $r < \sigma$ we still have Eq.~(\ref{eq1}); Eqs.~(\ref{eq6}),~(\ref{eq7}) fully account for the depletion
  attraction between the colloids caused by the polymers and show that the strength of this interaction can easily
  be controlled by variation of $\eta_p^r$. In the present study, we choose a single value of $\eta^r_p$ and a single
  choice of $q$ only,

  \begin{equation} \label{eq8}
  \eta_p^r =0.1, \quad q=0.15 \quad.
  \end{equation}

  We also note that for $\eta_p^r=0$ the model reduces to the simple hard sphere model, for which interfacial
  properties (see e.g. \cite{40,64} for references) and nucleation (e.g.~\cite{12,13}) have been studied extensively;
  but since there is evidence \cite{65} that hard spheres at hard walls (as well as on walls where a soft repulsion
  acts \cite{64}) exhibit complete wetting when the freezing fraction $\eta_f$ (cf. Fig.~\ref{fig1}) is approached,
  the simple hard sphere model is less suitable to study crystalline nuclei attached to flat walls, and shall not be
  considered here further.

  As already indicated in Fig.~\ref{fig2}, we choose a $L_x \times L_y \times D$ geometry with periodic boundary
  conditions in $x$ and $y$ directions, while soft repulsive walls occur at $z=0$ and $z=D$, respectively. These walls are
  described by a potential of the Weeks-Chandler-Andersen \cite{66} type

  \begin{eqnarray} \label{eq9}
  &&(k_BT)^{-1} V_{WCA} (z) = 4 \varepsilon[(\sigma_w/z)^{12} - (\sigma _ w / z)^6 + 1] \,\, \nonumber\\
  &&{\rm for} \, \, 0 \leq z \leq \sigma_w 2^{1/6} \quad,\nonumber\\
  && = 4 \varepsilon [(\sigma_w / ( D-z))^{12} - (\sigma_w/(D-z))^6 +1 ] \nonumber\\
  && {\rm for} \, (D-\sigma_w 2 ^{1/6})
  \leq z \leq D \quad ,\nonumber\\
  && = 0 \quad {\rm otherwise}.
  \end{eqnarray}

Here $\varepsilon$ describes the strength of the potential (in units of the thermal energy $k_BT$) and $\sigma_w$ its range;
in the present paper we only consider the case $\varepsilon=1$, $\sigma_w=\sigma/2$. Choosing $\sigma=1$ as
our unit of length, typical box linear dimensions were

\begin{equation} \label{eq10}
L_x=L_y=39.606548352, \quad D=40.313808144
\end{equation}

which means that for a typical packing fraction $\eta=0.511184$ those linear dimensions correspond to a total
number of colloids $N_c=61717$ in the system (allowing the observation of crystalline ``clusters'' containing
$4743$ colloids, for instance, see below).

\begin{figure} [ht]
\includegraphics[scale=0.35]{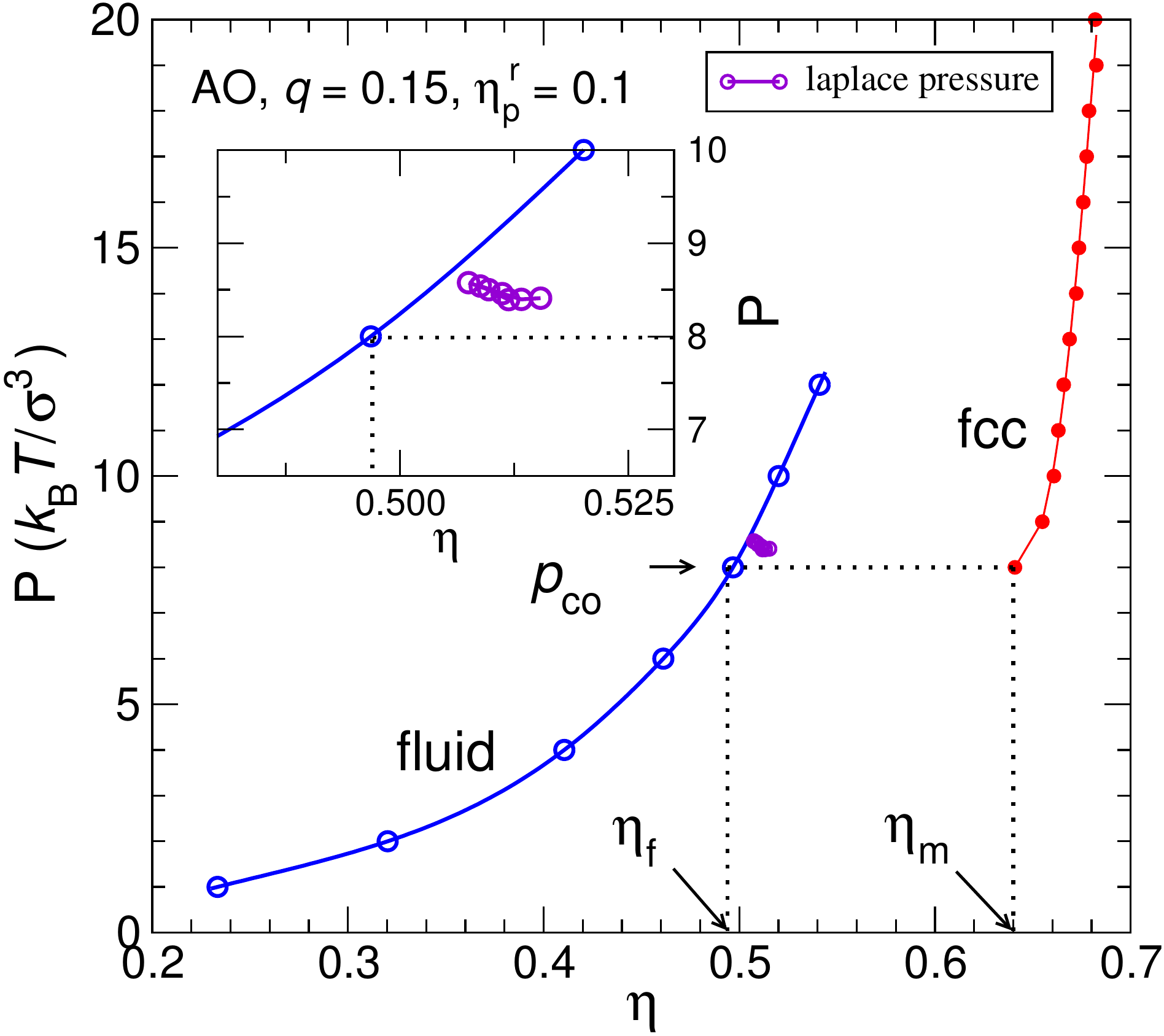}
\caption{\label{fig3} Pressure $p$ (in units of $k_BT/\sigma^3$) plotted vs. packing fraction, from the
AO model with $q=0.15$, $\eta_p^r=0.1$. This phase diagram was already obtained by Zykova-Timan et al. \cite{40}, who
found that fluid-solid coexistence occurs at $p_{co}=8.00$ (highlighted by the dotted line). For $\eta$
slightly larger than $\eta_f$, data points below the metastable bulk fluid branch (shown in the insert on
largely magnified scales) correspond to states of two-phase coexistence, where a solid wall-attached crystalline cluster
coexists with fluid phase, as shown schematically in Fig.~\ref{fig2}a. Note that $\eta_f=0.494$ and $\eta_m=0.64$,
respectively.}
\end{figure}

In addition to simulation boxes where the wall surface has square geometry, we have also chosen $L_x, L_y$ consistent
with a perfect triangular lattice (of the lattice spacing corresponding to $\eta_m$), to 
allow the formation of fcc crystalline layers with close-packed planes at the wall.

An important ingredient in our study is the accurate knowledge about the phase transition in the bulk for our model.
We have made the particular choice of Eq.~(\ref{eq8}) because for this choice the phase transition has already
been studied by Zykova-Timan et al.~\cite{40} using constant pressure Monte Carlo methods (NpT ensemble
\cite{67,68}). Fig.~\ref{fig3} shows the resulting equation of state: there are two branches of the pressure
versus packing fraction curve, a fluid branch and a solid branch, which corresponds to the face-centered
cubic (fcc) lattice structure. These data were obtained from simulations of cubic $L_x \times L_y \times L_z$ simulation
boxes with periodic boundary conditions, with $L_x=L_y=L_z=L$ for the fluid branch, while the ratios of $L_y/L_x$ and
$L_z/L_x$ for the simulation of the crystal were adjusted such that an integer number of close-packed lattice
planes stacked upon each other in the fcc ABCABC stacking sequence were compatible with the periodic boundary
conditions without distorting the lattice. As is evident from Fig.~\ref{fig3}, there occurs hysteresis over a
broad range of pressures. The estimation of the pressure $p_{co}$ at which phase coexistence occurs in equilibrium
was done \cite{40} using a method described in Refs.~\cite{40,69}. In short, in this method one prepares a slab configuration,
where in an elongated simulation box ($L \times L \times 5L)$ a crystalline domain is separated from liquid domains to the
right and to the left by flat domain walls running perpendicular to the $z$-direction. Precautions are taken to avoid
any elastic deformation of the crystal slab in such simulations. Then the average volume of the system $\langle V (t)
\rangle$ is determined in an $N P_z T$ Monte Carlo simulation as a function of Monte Carlo time for various pressures, 
and $p_{co}$ is found from the condition that $d \langle V(t) \rangle /
dt=0$, $t$ being the time variable in the simulation: if $p < p_{co}$, the crystal shrinks on average (and $V$ increases), while for
$p > p_{co}$ the crystal grows. This method
has been carefully tested for the simple hard sphere model \cite{40,69}, and found to give very accurate results;
we expect it to work for the AO model similarly well.

\begin{figure} [ht]
\includegraphics[scale=0.350]{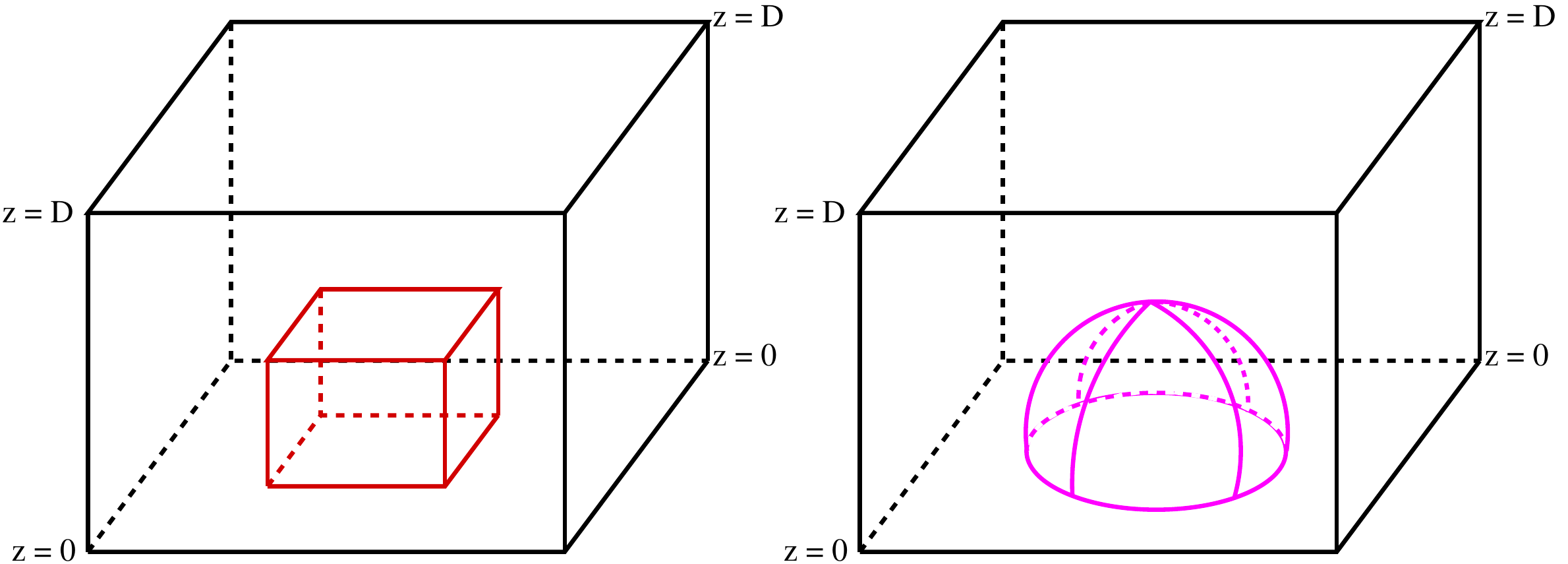}
\caption{\label{fig4} Initialization of the system: To realize a single crystalline cluster attached to the wall at
$z=0$, we prepare the total system first at packing fraction $\eta_f$, and then replace either a cuboid (left)
or a hemisphere (right) by a face centered cubic crystal, with the close packed (111) planes stacked parallel to the
plane $z=0$. The volume of the cuboid (or hemisphere) is chosen in accord with Eq.~(\ref{eq13}) at each chosen value of $\eta$.} 
\end{figure}

As a complementary approach to obtain the equation of state from simulations in the NpT ensemble, we have developed a method
to very accurately compute the pressure from simulations in the constant volume (NVT) ensemble \cite{70}. While
the contribution to the pressure from the attractive part of the potential \{Eq.~(\ref{eq6})\} is straightforwardly obtained
from the standard virial expression \cite{67,68}, care is needed for the accurate estimation of the pressure contribution due to the hard core repulsion, Eq.~(\ref{eq1}) \cite{70}. In a recent paper, we have shown that this task can be solved by
applying the method due to DeMiguel and Jackson \cite{71}. In this way one can obtain both the hydrostatic pressure of a
bulk system as a function of the packing fraction, and for a system with external walls one can extract the wall tension as
usual \cite{16} from the anisotropy of the pressure tensor. We refer the interested reader to Refs. \cite{64,70} for details
on this method. Here, we only emphasize the two following facts: (i) for a homogeneous bulk system, the function
$p(\eta)$ found in the NVT ensemble precisely coincides with its counterpart in the NpT ensemble \cite{70}: thus finite size
effects due to the change of the statistical ensemble are negligibly small in our problem. (ii) The method of Ref. \cite{70} yields also the local transverse component $P_T(z)$ if we deal with a system confined by walls. This allows to estimate the pressure
$p_f(\eta)$ in the fluid part of a system that has separated into a solid cluster and surrounding fluid. Since this fluid
coexisting with a crystalline cluster of finite size does not have the coexistence pressure $p_{co}$ of the bulk, but rather the pressure
of the fluid coexisting with the crystalline cluster is enhanced (Laplace pressure), estimation of this pressure is nontrivial and of
interest: these data hence are included in Fig.~\ref{fig3}, but we defer their analysis to the next section.

At this point, we emphasize that it is rather straight-forward to prepare (Appendix A) a system at packing fraction near $(\eta_f + \eta_m)/2$, the center of the two-phase coexistence region, then the system develops easily towards an equilibrium state, where both walls are coated by thin crystalline films (Fig.~\ref{fig2}c) with a liquid state of appropriate thickness in between, and the pressure inside the liquid must be the coexistence pressure $p_{co}$. By this method, the results of \cite{40} could be checked independently.
 
Note that special care is needed to prepare a system where a single crystalline cluster coexists with surrounding fluid
with which it is in equilibrium. When we would start out with a packing fraction $\eta$ that exceeds $\eta_f$ slightly, such
as those values from which data in Fig.~\ref{fig3} are included, but choose an initial state that is similar to the
bulk metastable fluid (without walls) as an initial condition, the system develops towards a metastable state where
the density of colloids just show the familiar layering at both walls (cf. the analogous data for hard spheres at $\eta <
\eta_f$ presented in \cite{64}), but it would take an unacceptable long time in the simulation until a wall-attached solid cluster
would be nucleated. This expectation actually is born out by the simulations (see next section). This fact is understandable from
rough estimates of the barriers for heterogeneous nucleation, obtained from the standard classical theory \cite{9,14,15} for the
sizes of solid clusters as studied here, which need only the contact angle $\theta$ (which is estimated independently for our
system, see below) as an input: according to the classical Turnbull \cite{9,10} theory of heterogeneous nucleation the volume
$V^*$ of a critical droplet having a sphere cap shape and contact angle $\theta$ (cf. Fig.~\ref{fig4}), the associated
free energy barrier is

\begin{equation} \label{eq11}
V^*=(4 \pi R^{*3}/3) f (\theta), \quad \Delta F^*_{\rm het} = \Delta F^*_{\rm hom} f(\theta)
\end{equation}

where

\begin{equation} \label{eq12}
\Delta F^*_{\rm hom} = \frac{4 \pi}{3} R^{*2} \gamma_{sl} \quad , \quad f(\theta)=(1-\cos \theta)^2
(2 + \cos \theta )/4 \quad .
\end{equation}

In Eqs.~(\ref{eq11}),~(\ref{eq12}) it is assumed that the dependence of the solid-liquid interface tension
$\gamma_{sl}$ on interface orientation can be neglected, and it is also assumed that $\gamma_{sl}$ does not
depend on the radius of curvature $R^*$ of the ``droplet''. Then the critical droplet that forms in a homogeneous
nucleation process has a spherical shape, and the free energy barrier against homogeneous nucleation
$\Delta F^*_{\rm hom}$ is just (1/3) of the total surface free energy, $4 \pi R^{*2} \gamma_{sl}$ of the
critical droplet \cite{1,2}. For heterogeneous nucleation, the free energy barrier that needs to be
overcome, is reduced by the same factor $f(\theta)$ as the volume of the sphere cap is reduced in comparison
with the full volume of the sphere \cite{9,10}.

Estimates of the contact angle from Young's equation (see below) imply that $\theta$ is at least as large
as $\theta \approx 70^o$, and hence we can conclude that $f(\theta)= 5/32$ (or larger). Using then a
particle number $N^*=4413$ in our solid cluster (which is a typical example) and taking for the solid the packing
fraction at coexistence pressure, $\eta_m=0.64$, we find that the corresponding volume is $V^*=N^*/
\rho=N^* \pi \sigma^3/(6 \eta_m) \approx 3610$ (remember that $\sigma=1$ is our unit of length). Using
then the estimate for $f(\theta)$ as quoted above we find $R^*\approx17.67$ and $\Delta F^*_{\rm het} \approx2043 \gamma_{sl}$.
Now the estimate of Zykova-Timan et al. \cite{40} for our model \{Eq.~(\ref{eq8})\} is $\gamma_{sl} \approx (0.95 \pm 0.05)k_BT/\sigma^2$. Consequently, one would predict a free energy barrier as large as $\Delta F^*_{\rm het} \approx 2000 k_BT$!
Even if this estimate would be an overestimate by a factor of two or three (which is well possible in view of the crudeness
of the approximations Eqs.~(\ref{eq11}),~(\ref{eq12}) for crystal nucleation) still the spontaneous formation of such large
crystalline clusters as studied here would never be visible in a simulation.

Thus the recipe to study large wall-attached crystalline clusters is to prepare the system in an initial
state from which there is either only a low free energy barrier to be crossed for the system on its way toward
thermal equilibrium (or, even better, no barrier at all). This is achieved by putting a crystalline seed of
roughly the right size into the box (Fig.~\ref{fig4}). This seed either has the shape of a cuboid or hemisphere.
In either case an integer number of close-packed (111) lattice planes of the fcc structure are stacked upon
each other parallel to the confining wall at the bottom of our ``container''. The lattice constant of this
crystalline cluster is chosen such that the crystal has a packing fraction $\eta=\eta_m$. The volume
of the crystal is cut out from a $L \times L \times D$ simulation box filled by well equilibrated fluid
at packing fraction $\eta_f$. The lever rule then fixes the volume of the crystal nucleus for a
given choice of $\eta$ for $\eta_f < \eta < \eta_m$:

\begin{equation} \label{eq13}
V\eta=V_{\rm crystal} \eta_m + (V-V_{\rm crystal}) \eta_f
\quad .
\end{equation}

Of course, Eq.~(\ref{eq13}) disregards finite size effects: the packing fraction of the liquid coexisting with a
nanoscopically small crystal is expected to slightly exceed $\eta_f$ because the density of the fluid surrounding 
the crystalline cluster must be enhanced due to the Laplace pressure associated with a small ``droplet''; similarly, also the density
of the nanocrystal may differ somewhat from its macroscopic counterpart. However, for the already somewhat large
($N^* > 1000)$ nanocrystals studied here, these effects did not prevent the successful equilibration of the crystalline clusters.
As a first step of this equilibration, forbidden overlaps of particles in the crystal and in the fluid are removed.
We found that using a hemisphere as an initial state equilibrium is reached more rapidly than with a cuboid as initial
state; but the properties of the equilibrium that is reached do not depend on the initial state, as it should be. Of
course, in equilibrium we could have the solid cluster attached to the wall at $z=D$ with the same probability as the
situation that is actually studied: Due to the initialization (Fig.~\ref{fig4}) the symmetry between both otherwise
identical walls is broken ``by hand''. Note also that a consideration along the lines of Eqs.~(\ref{eq11}),~(\ref{eq13}) readily
shows that a situation with two solid clusters (one cluster at each of the walls at $z=0$ and $z=D$) is less favorable
than the single cluster state.

In order to study the detailed physical properties of the crystalline cluster, it is necessary to identify in microstates
of the system which particles belong to the solid and which particles belong to the liquid. For this purpose, we follow
the traditional methods \cite{40,72,73} where the ``coherence property'' of a particle and its nearest neighbors are
analyzed using spherical harmonics. Specifically, one computes the complex vector $\vec{q}_6 (i)$ for each particle (labeled
by index $i$). The 13 components (labeled by $m$) of this vector depend on the relative orientation of the ``bond'' connecting
the particle to its neighbors, and are defined as \Big\{$\tilde{Q}_{6m} (i)= (1/N_b(i)) \sum\limits_{k=1}^{N_b(i)} Y_{6m}^{(k)}\Big\}$

\begin{equation} \label{eq14}
q_{6,m} (i) =\frac{\tilde{Q}_{6m} (i)}{(\sum\limits_{m-6}^{+ 6} |\tilde{Q}_{6m} (i) |^2)^{1/2}}
\end{equation}

where $N_b (i)$ is the number of nearest neighbors of particle $i$, $k$ labels the bonds connecting particle $i$
with its $k$'th neighbor, and $Q_{lm}$ are spherical harmonics ($l=6$ has to be used in the present case). Such nearest
neighbors of the $i$'th particle are identified by defining a cutoff distance, and all the particles whose relative distance
from the $i$'th particle is within the cutoff range are identified as candidates for being nearest neighbors. 
The cutoff distance is chosen as the first minimum in the radial distribution function of the colloid particles. 
Then we compute $d_6(i)$ according to

\begin{equation} \label{eq15}
d_6(i) = \sum\limits_{k=1}^{N_b(i)} \sum\limits_{m=-6}^{+6} q_{6,m}(i) \cdot q^*_{6m} (k) \quad .
\end{equation}

We consider $i$ as a particle belonging to the solid if $d_6(i) \geq 0.85$ and its number of nearest
neighbors is $N_b(i) \geq 12$. The latter choice has the consequence that the interface between the crystal and the liquid 
is put towards the crystal region of the (extended \cite{40}) liquid-crystal interfacial profile, rather than into its center, so some roughness from the surface of the crystal is eliminated, and the number of particles in the crystal slightly underestimated. But we expect that this choice will not lead to noticeable systematic errors of the contact angle of the crystalline cluster. If we choose a 
smaller value than 12 for this cutoff, also small clusters in the liquid, which are not of physical significance, are counted as being crystalline.

\section{RESULTS AND DISCUSSION}

\begin{figure} [ht]
\includegraphics[width=0.23\textwidth]{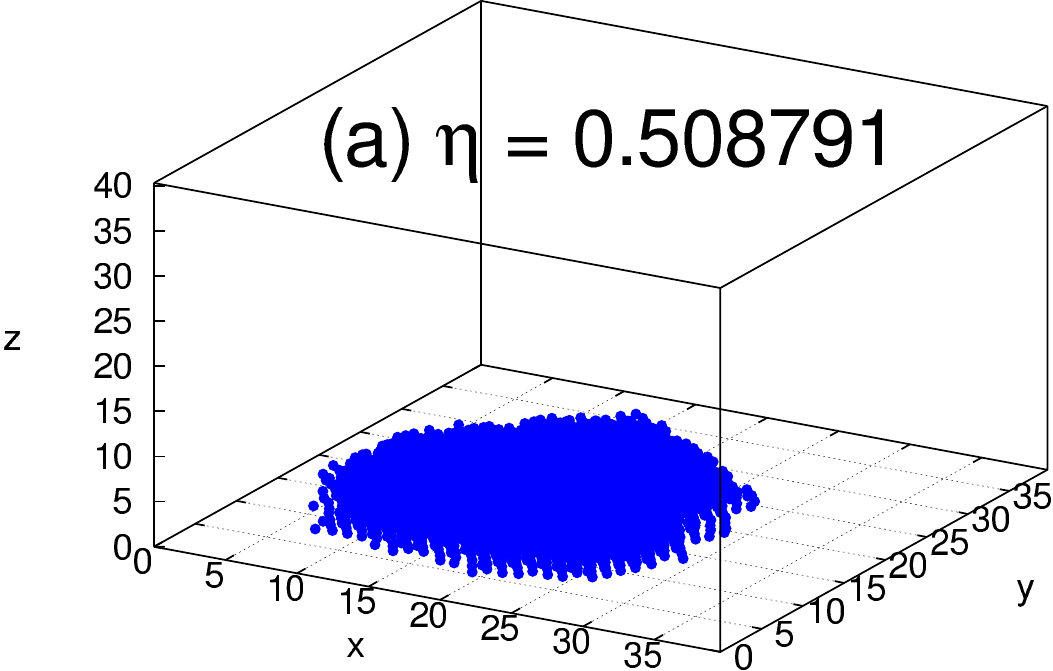} 
\includegraphics[width=0.23\textwidth]{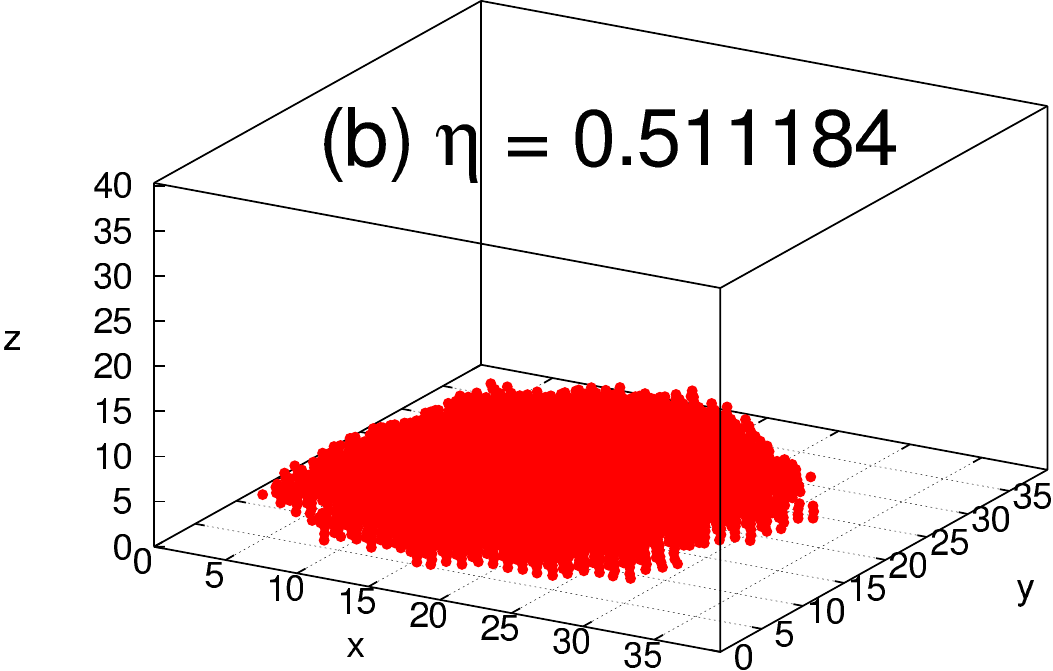}
\includegraphics[width=0.23\textwidth]{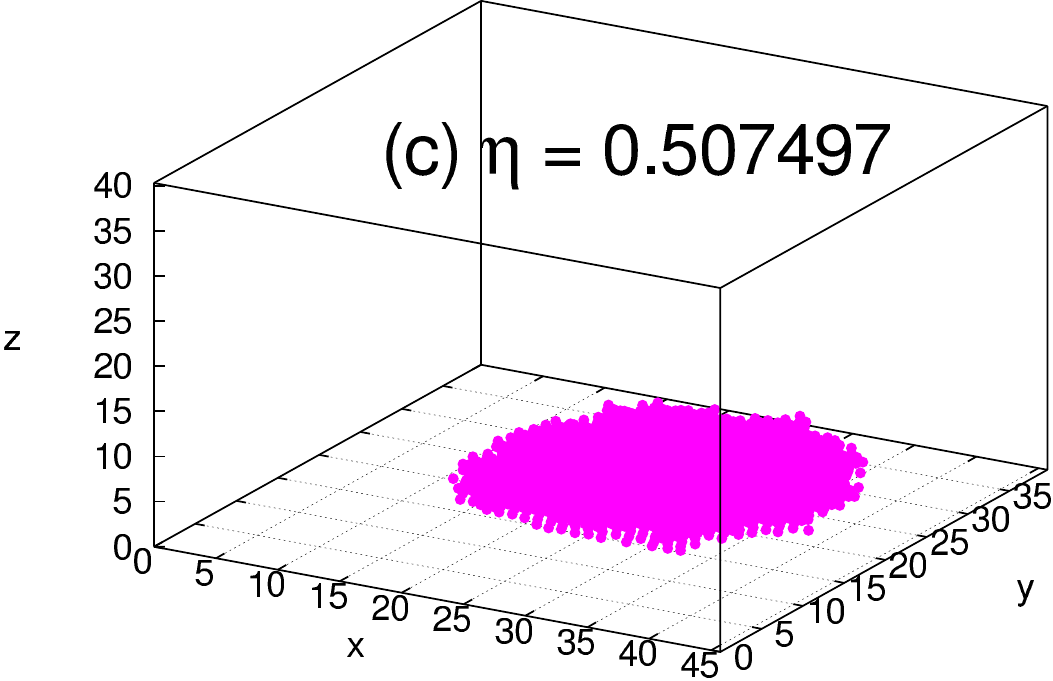}
\includegraphics[width=0.23\textwidth]{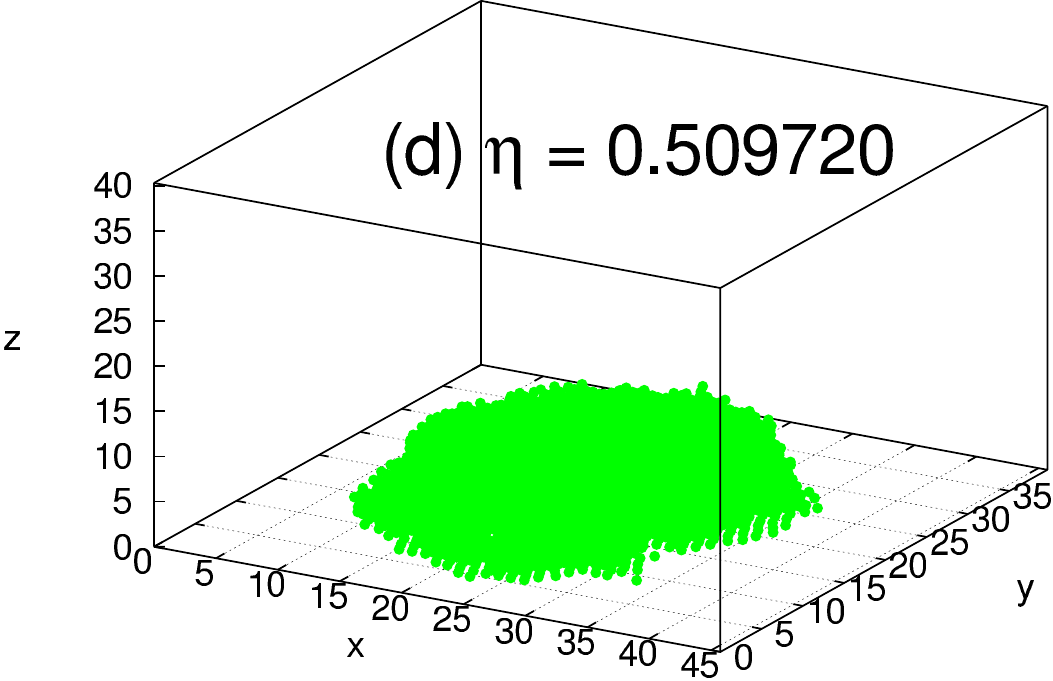}
\caption{\label{fig5} Examples of typical crystalline clusters, as seen from above in a three-dimensional
view. The number of solid particles in the clusters are $N^*$=3092 (a), 4743 (b), 2797 (c) and 4852 (d). Size of the system used for (a)-(b) $L_x=Ly=39.60654835212$, $D=40.31380814413$ and that for (c)-(d) is $L_x=45.73370270554$, $L_y=\sqrt{3}/2L_x$, $D=40.31380814413$. Corresponding total volume fractions in the $L_x \times L_y \times D$ simulation box are quoted in the figure. All crystals shown here have been chosen with their (111) planes parallel to the substrate.}
\end{figure}

\begin{figure} [ht]
\includegraphics[width=0.2\textwidth]{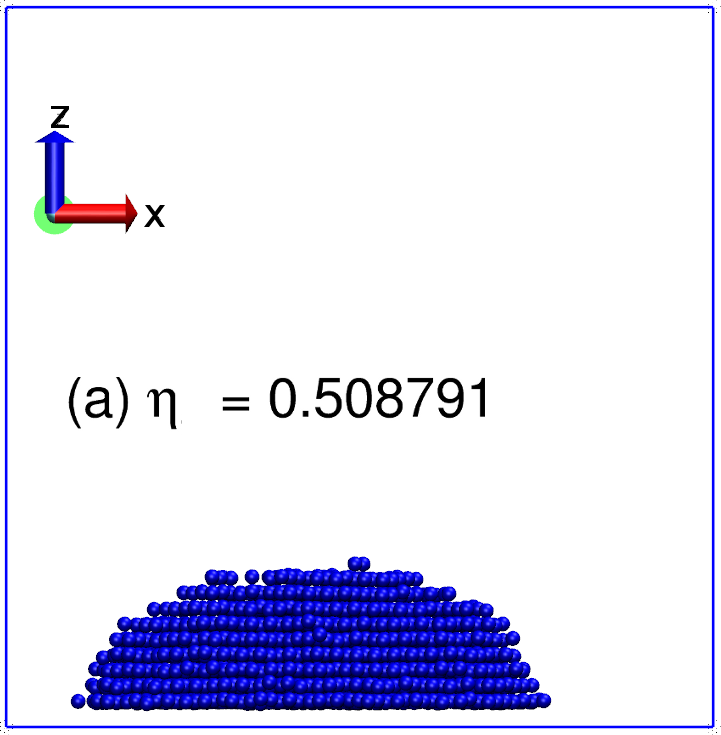} 
\includegraphics[width=0.2\textwidth]{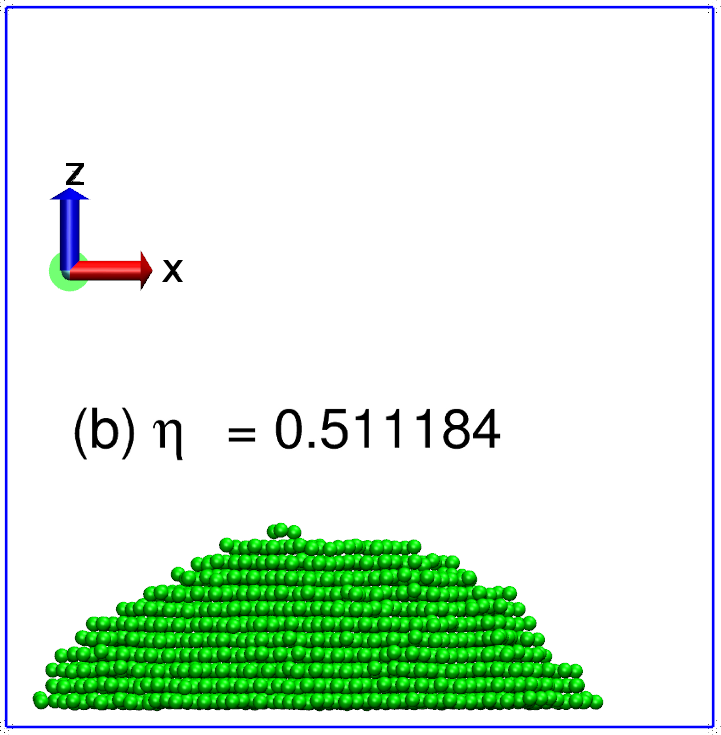}
\includegraphics[width=0.2\textwidth]{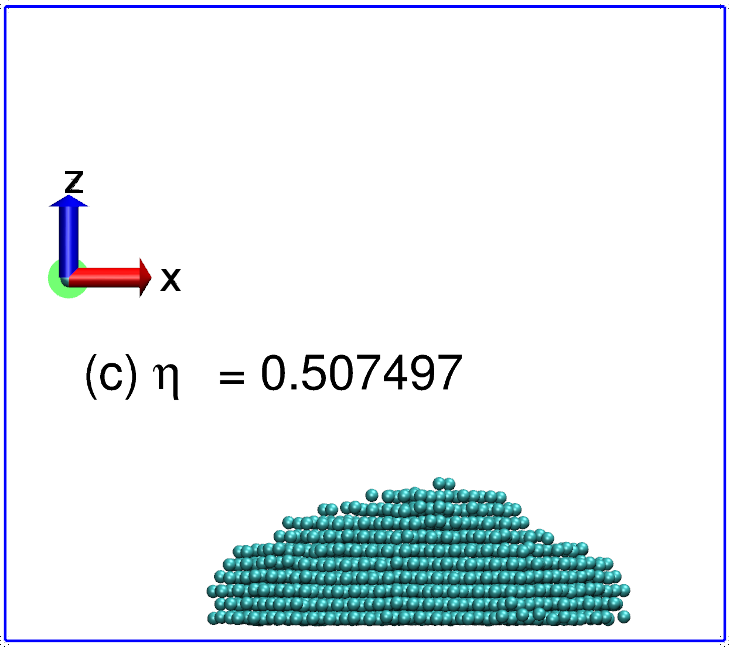}
\includegraphics[width=0.2\textwidth]{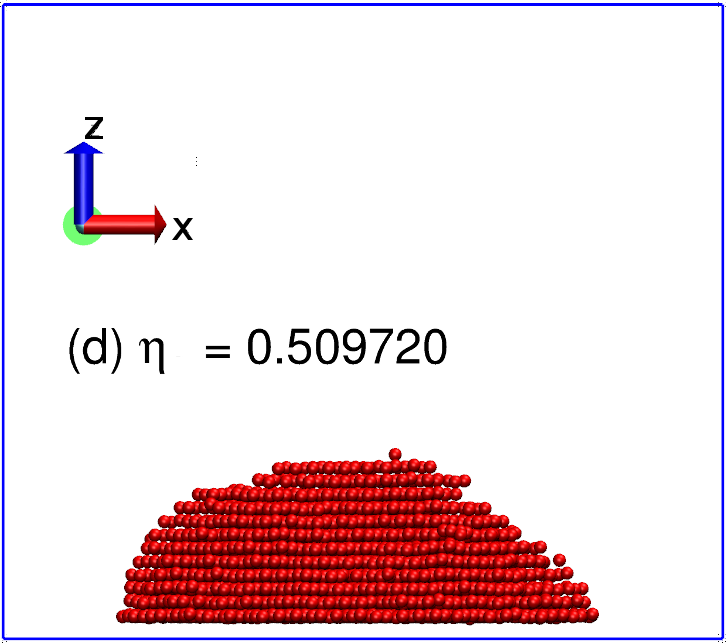}
\caption{\label{fig6} Same as Fig.~\ref{fig5}, but viewing the projection into the $xz$-plane.}
\end{figure}

\begin{figure} [ht]
\includegraphics[width=0.23\textwidth]{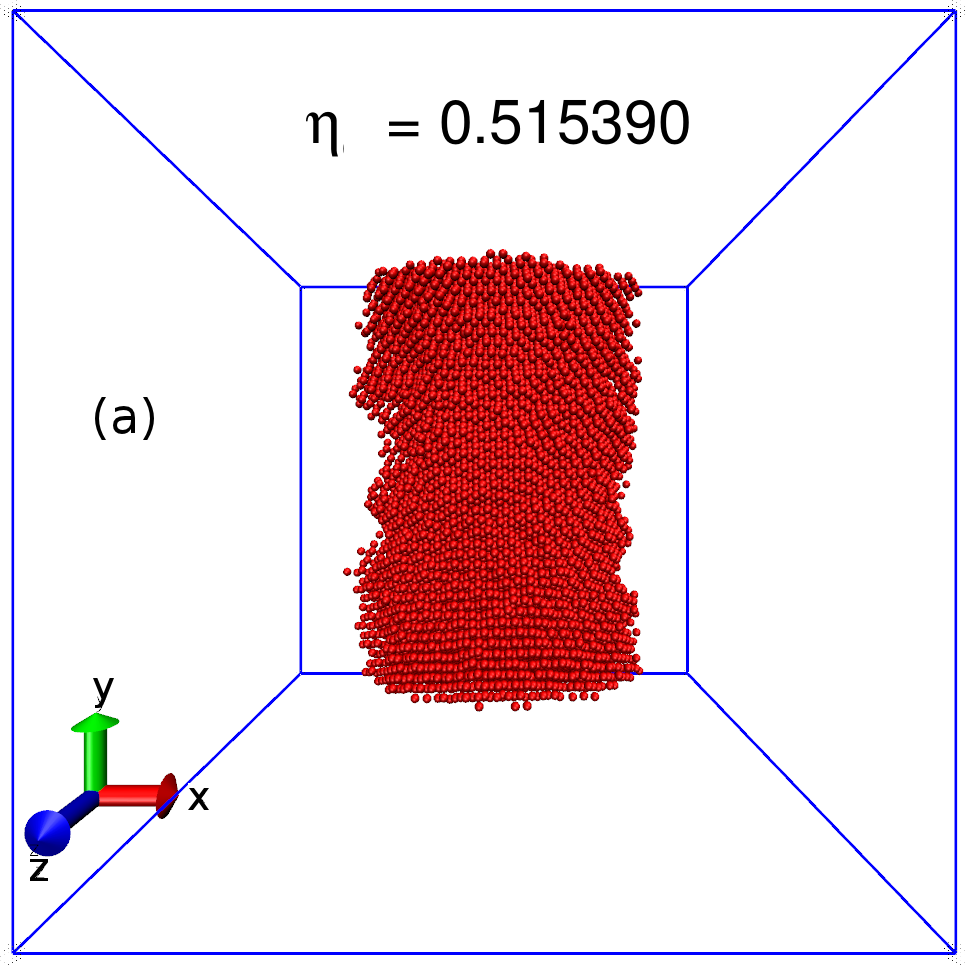} 
\includegraphics[width=0.23\textwidth]{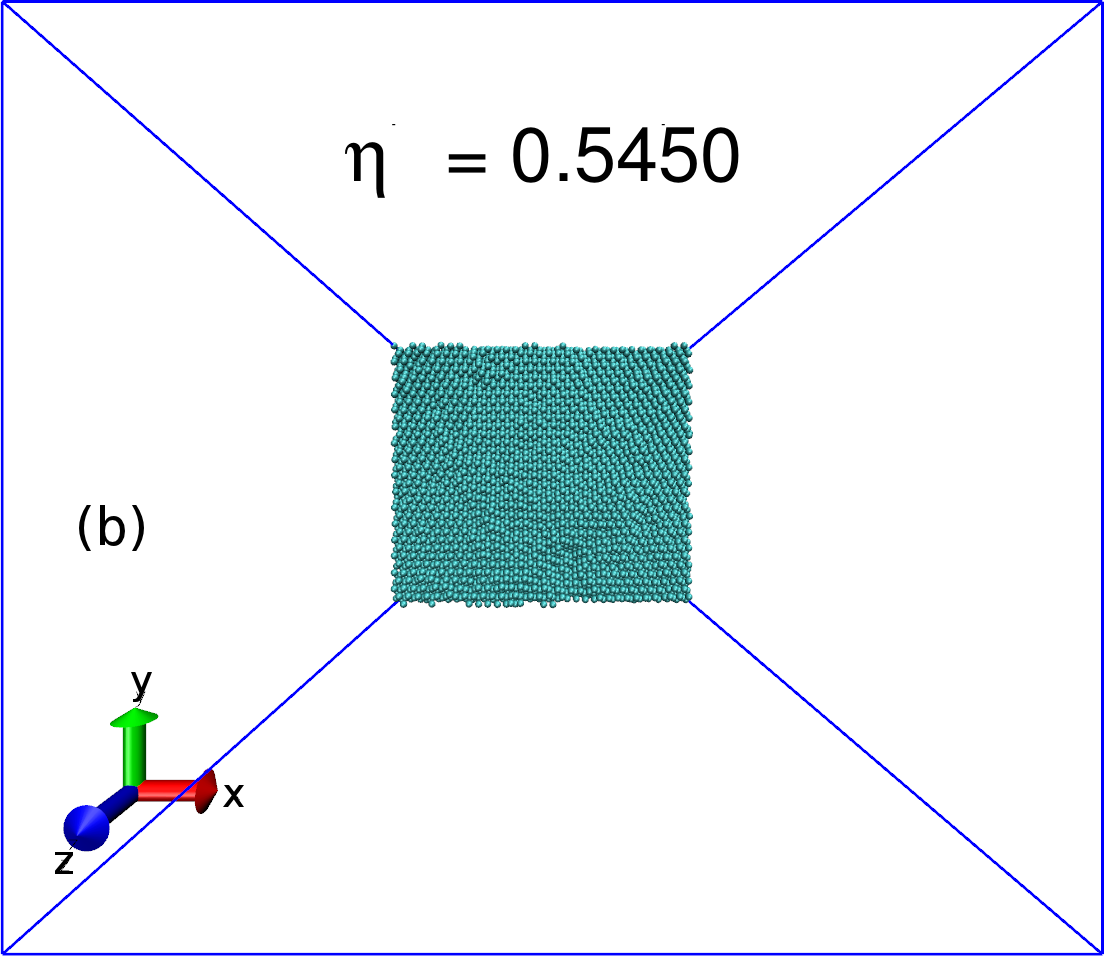}
\caption{\label{fig7} Same as Fig.~\ref{fig5}, but for $\eta=0.51539(a)$ and $\eta=0.545(b)$. In case (a), a (distorted) cylindrical domain has formed, while (b) corresponds to a slab-like configuration. The number of solid particles in the clusters are $N^*=$ 8358 (a) and 23823 (b). Size of the system used for (a) is $L_x=L_y=39.60654835212$, $D=40.31380814413$, and that for (b) is $L_x=33.59395945368$, $L_y=\sqrt{3}/2L_x$, $D=65.42781457594$.}
\end{figure}

Figs.~\ref{fig5},~\ref{fig6} show typical snapshot pictures of the crystalline clusters obtained by the
method as described above (particles identified as fluid are not shown). One can see that we do obtain
crystalline clusters of roughly sphere cap shape (note, however, that there occur substantial fluctuations
in both the size and the shape of these clusters, as expected, since we do not apply any constraint to the
properties of the solid cluster, other than that the lever rule, Eq.~(\ref{eq13}), must be satisfied,
since the total particle number in the simulation box is a conserved quantity). Fig.~\ref{fig7} shows cases where
$\eta$ was chosen too large (for the considered choice of box linear dimensions) so that no longer a single
sphere cap is stable, but rather the system forms a (distorted) cylindrical cluster (connected in itself by the
periodic boundary condition, as drawn schematically in Fig.~\ref{fig2}) or even a slab-like configuration forms.

\begin{figure} [ht]
\includegraphics[scale=0.350]{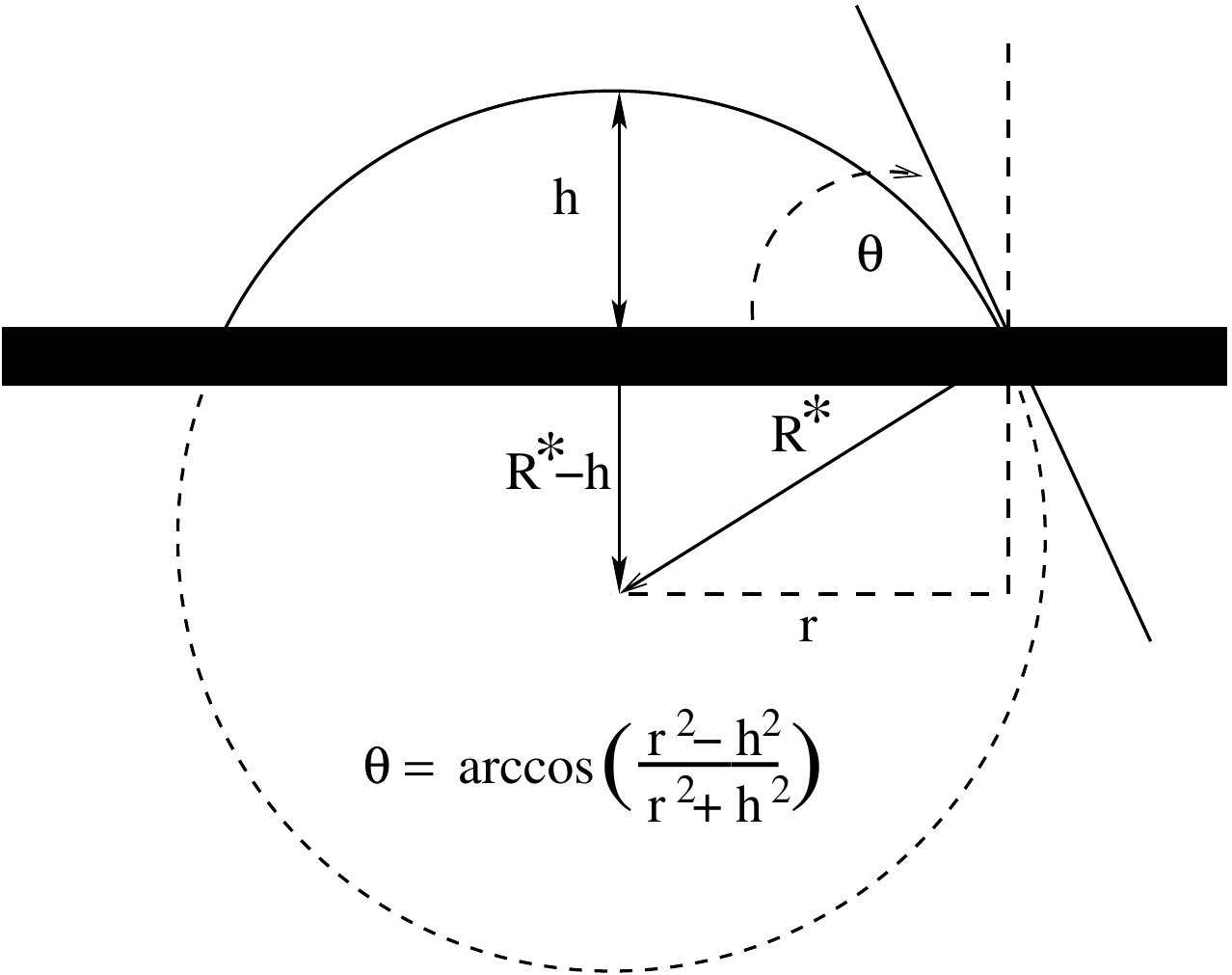}
\caption{\label{fig8} Measurement of the contact angle of a crystalline cluster assuming a sphere cap shape. When a crystalline cluster is identified the number of particles $N_b$ in the first crystalline layer adjacent to the basal plane of the substrate is estimated. From the area $N_b \sigma^2$ of this layer we obtained the radius $r$ in the basal plane as $r= \sqrt{N_b/\rho_A\pi}$ (remember $\sigma=1$), here $\rho_A$ is the areal density of the cluster base which is equal to $\rho_A=\frac{4}{\sqrt{3}}(\frac{6\eta_m}{4\pi})^{2/3}$. From the total number $N^*$ of particles in the crystalline cluster one then obtains the height $h$ of the sphere cap via $N^*=\eta_m h(3r^2 + h^2)$. The contact angle then follows as $\theta= \arccos[(r^2-h^2)/(r^2 + h^2)]$.}
 \end{figure}

\begin{figure} [ht]
\includegraphics[scale=0.312]{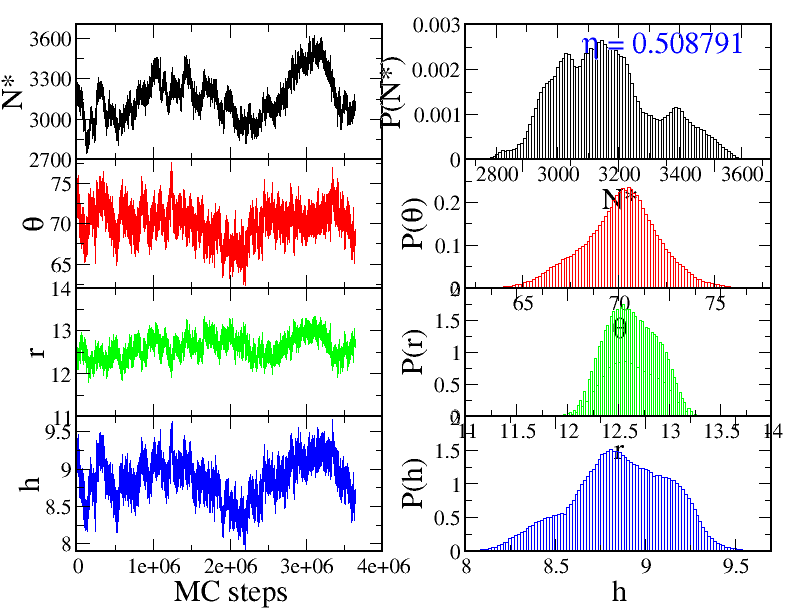}
\caption{\label{fig9} Properties of wall-attached crystal clusters, for a packing fraction $\eta=0.508791$ of particles in the simulation box ($L_x=L_y=39.60654835212$ and $D=40.31380814413$). Left panel shows the time variation of the cluster size $(N^*)$, contact
angle $(\theta)$, basal radius $(r)$ and height $(h)$. The right panel shows the resulting distribution functions of these quantities.}
\end{figure}

\begin{figure} [ht]
\includegraphics[scale=0.312]{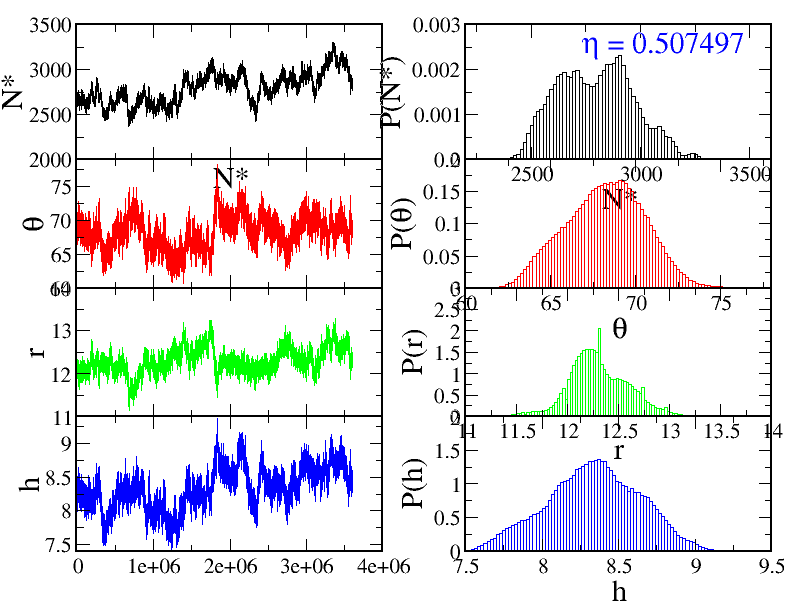}
\caption{\label{fig10} Same as Fig.~\ref{fig9}, but for $\eta=0.507497$ ($L_x=45.73370270554$, $L_y=\sqrt{3}/2L_x$ and $D=40.31380814413$).}
\end{figure}

\begin{figure} [ht]
\includegraphics[scale=0.312]{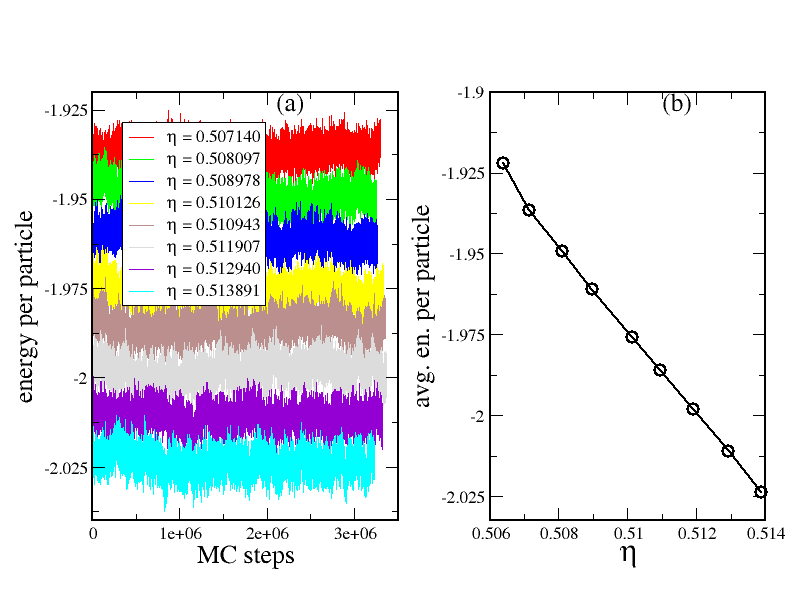}
\caption{\label{fig11} (a) Time evolution of the total internal energy per particle for several choices of packing fraction $\eta$, as indicated and (b) the average total internal energy per particle plotted vs. packing fraction. Size of the system used for all the choices of packing fraction is $L_x=L_y=45.26462668814$ and $D=40.31380814413$. }
\end{figure}

From Figs.~\ref{fig5} -~\ref{fig7} it is clear, that reliable data on crystalline clusters can only be obtained
as long as the lateral linear dimensions of the crystalline cluster are distinctly smaller than the box linear
dimensions $L_x$ and $L_y$. We shall disregard conditions where cylinder-like and slab-like domains form
in the following. The solid clusters that are not affected by the lateral periodic boundary conditions are analyzed
under the assumption that a sphere-cap shape is a reasonable approximation (Fig.~\ref{fig8}).
Figs.~(\ref{fig9}),~(\ref{fig10}) show typical results for the time evolution of the cluster size $N^*$,
contact angle $\theta$, basal radius $r$ and cluster height $h$, and the resulting probability distributions of
these quantities. Despite the use of a significant computational effort (typically a million Monte Carlo steps per
particle, for systems containing on the order of 60000 particles, were used) the amplitude of the fluctuations
in the crystalline cluster properties are large, and the correlation time of these fluctuations typically is of the order
of 10$^5$ MCS. Thus it is difficult to ascertain the systematic trend that the comparison of Figs.~\ref{fig9}
and~\ref{fig10} suggests, namely that there is a systematic increase of the contact angle with the size of the
crystalline cluster. Note that we have made runs both for a square basal plane of the box ($L_x=L_y$) and for a
hexagonal base, but we did not find that this choice leads to systematic differences. We also emphasize that the
total internal energy in the system is much less fluctuating, and its average shows a smooth variation with
$\eta$ (Fig.~\ref{fig11}). We have taken all precautions that our systems are well equilibrated and our runs
do constitute a significant statistical effort: but the equilibrium between the (sphere-cap shaped) solid
cluster and its environment, which is only stabilized by the constraint of constant density \{Eq.~(\ref{eq13})\}
and which would not be stable in the finite box if one could carry out the simulation at constant chemical
potential rather than at constant density, allows very strong and long-lived fluctuations. There is no constraint on
the shape of the crystalline cluster other than the driving force to minimize the free energy of the total system. The same
fact holds concerning the size of the cluster: a large fluctuation increasing the cluster size needs a density
fluctuation in the surrounding fluid which then has a too low density. But the driving force to bring
the fluid density back to its appropriate value is rather weak, and hence it may need a long time for the
cluster size to return to its equilibrium value. Similar considerations apply to other cluster properties as well. In any
case, these large fluctuations of the contact angle, height, basal area, and volume of the cluster constitute
evidence that the cluster surface is rough, rather than faceted: in the latter case much less fluctuations
would be expected.

Given the fact that from the averaging of properties of the observed wall-attached crystalline clusters (Fig.~\ref{fig9},~
\ref{fig10}) one can conclude that the concept of wall-attached sphere-cap shaped ``droplets'' is at least qualitatively
reasonable as a coarse-grained description, it makes sense to compare the observed contact angle of the ``droplets''
with the ``macroscopic'' contact angle. The latter results from Young's equation \cite{16,17,18} as

\begin{equation} \label{eq16}
\gamma_{wf} - \gamma_{wc} = \gamma_{fc} \cos \theta \quad .
\end{equation}

Here $\gamma_{wf}$ is the excess free energy of the fluid due to the confining wall, and $\gamma_{wc}$ the
analogous quantity of the crystal, for the case that the close-packed planes in the fcc crystal are parallel to
the flat wall surface (consistent with what is actually observed, Fig.~\ref{fig6}). Note that in Eq.~(\ref{eq16})
it is explicitly assumed that the interfacial tension of the fluid-crystal interface does not depend on the orientation
of this interface, which clearly is an approximation, and will not hold true in general. If we nevertheless accept this
approximation, we can conclude from the work of Zykova-Timan et al, \cite{40} that $\gamma_{fc}=(0.96 \pm 0.05)k_BT/\sigma^2$.
For the estimation of the wall tensions $\gamma_{wf}$ and $\gamma_{wc}$ recently several fairly accurate methods
were developed \cite{64,70}.

\begin{figure} [ht]
\includegraphics[scale=0.375]{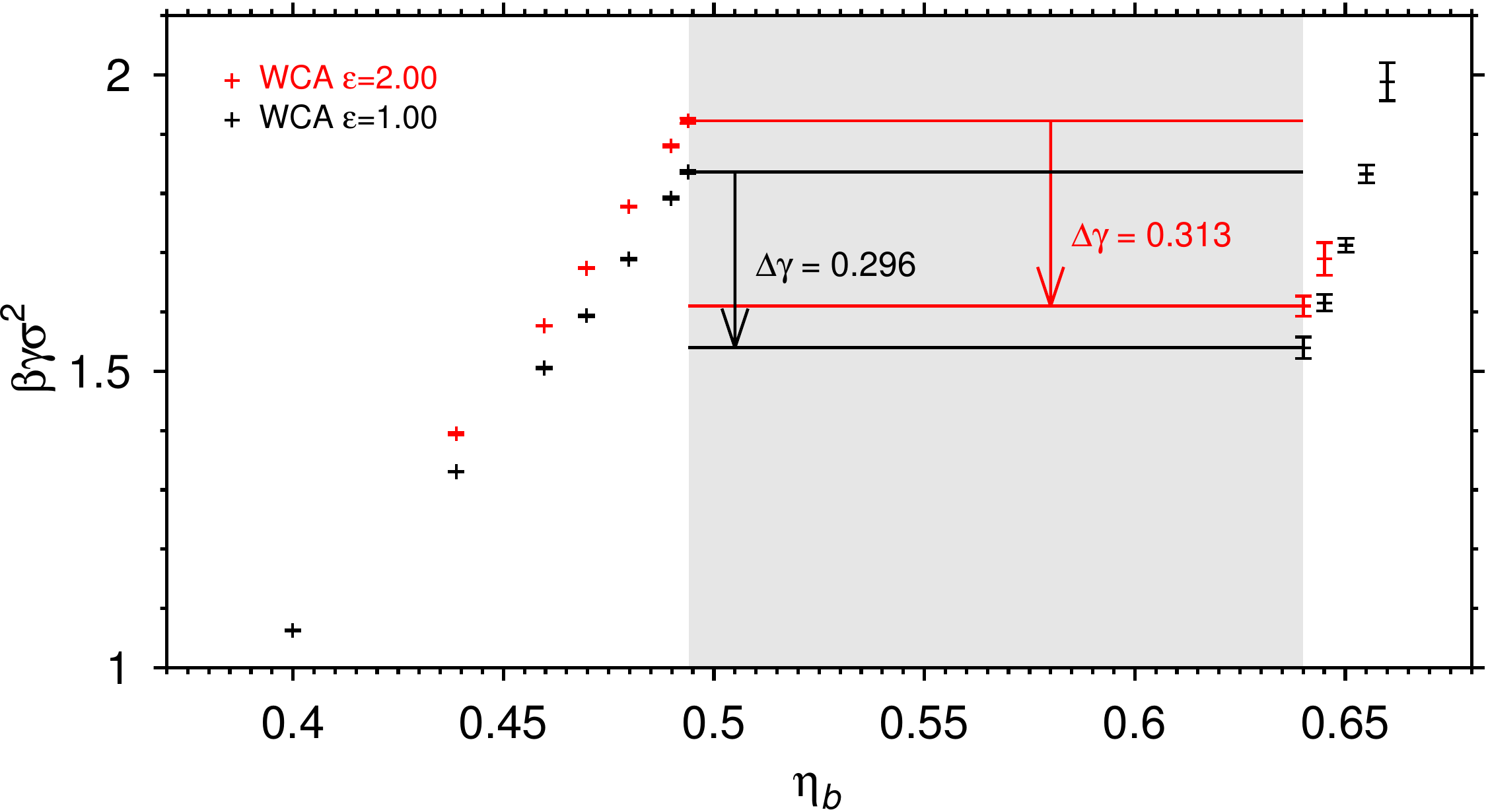}
\caption{\label{fig12} Wall tension of the AO model plotted vs. bulk colloid packing fraction $\eta_b$, as obtained from the ``ensemble mixing method'', as described in \cite{64,70}. Two choices of WCA potential are included, and the resulting differences $\Delta \gamma= \gamma_{wf} - \gamma_{wc}$ at the phase transition are quoted. Note that the horizontal straight lines indicate the two-phase coexistence region.}
\end{figure}

 Fig.~\ref{fig12} plots results for the wall tensions $\gamma_{wf}$, $\gamma_{wc}$ for the present
 model (defined by Eqs.~(\ref{eq6}),-~(\ref{eq9})). Using then the resulting estimates $\Delta \gamma=\gamma_{wf}-
 \gamma_{wc}$ at the transition we can use Eq.~(\ref{eq16}) to predict the contact angle $\theta$, which turns
 out to be close to 70$^o$. Fig.~\ref{fig12} shows that $\Delta \gamma$ (and hence $\theta$) depend on the
 strength $\varepsilon$ of the WCA potential only rather weakly. A related finding was already reported
 in \cite{64} for the simple hard sphere fluid (for which complete wetting seems to occur; i.e.~
 $\theta \approx 0^o$). These results imply that the variation of the strength of the inter-particle attraction
 ($\eta_p^r$) is a suitable recipe to change the contact angle of the system, while the variation of the
 strength of the wall repulsion $(\varepsilon)$ is not. Recall that experimentally $\eta^r_p$ can be varied
 by changing the polymer concentration in the system, while $\varepsilon$ could be varied by different wall
 coatings (e.g., using a polymer brush layer of variable grafting density).
 
\begin{figure} [ht]
\includegraphics[scale=0.313]{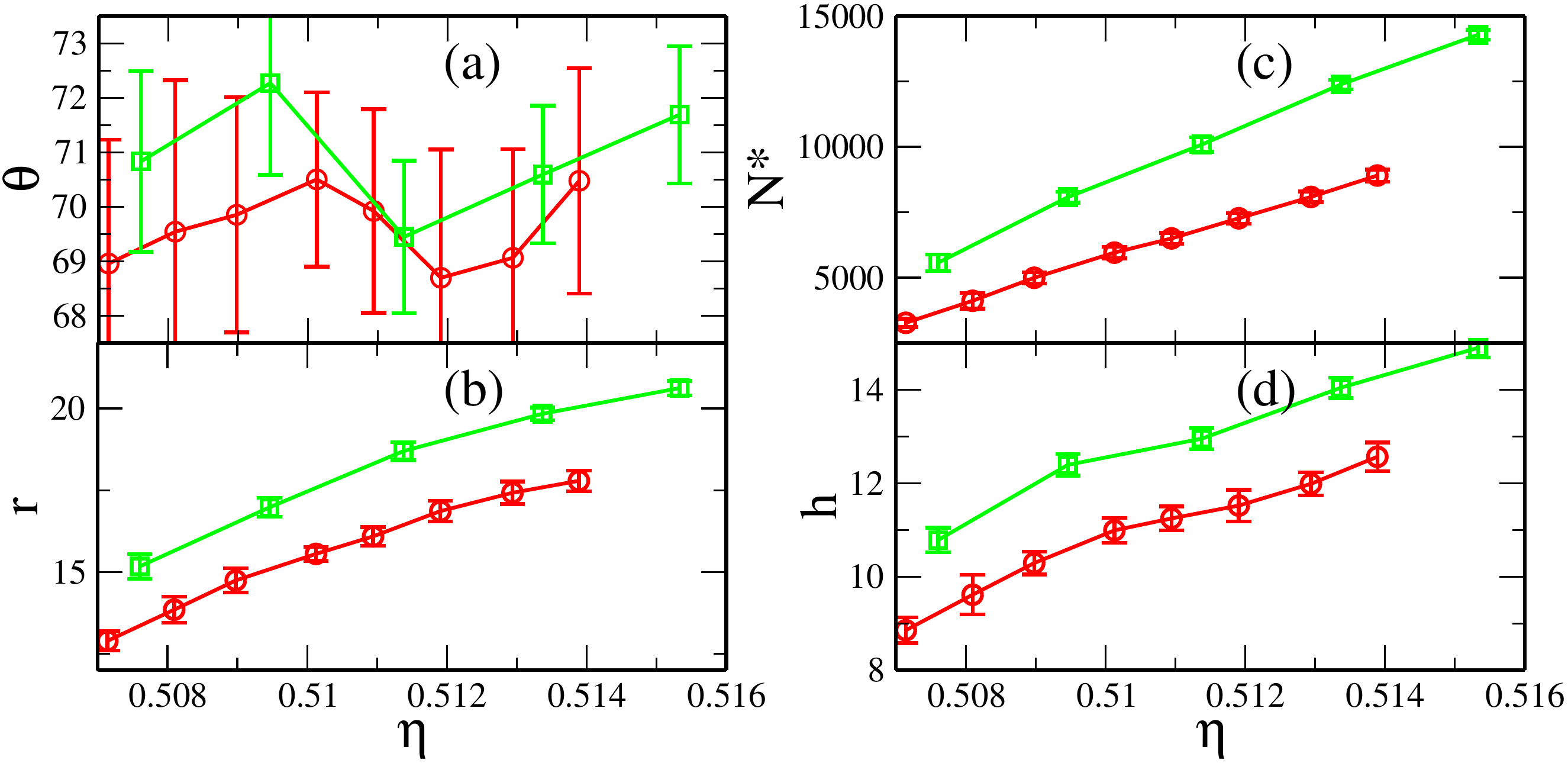}
\caption{\label{fig13} Properties of wall-attached crystalline clusters, shown as functions of the packing fraction $\eta$: contact angle $\theta$ (a), basal radius r (b), particle number N* in the crystalline cluster (c) and height h (d). Two choices of total box size are included: the smaller system has $L_x = L_y = 45.26462668814, D = 40.31380814413$ (red color data sets with $\bigcirc$ symbol), the larger system has $L_x = L_y = 50.92270502416, D = 44.55736689614$ (green color data sets with $\Box$ symbol).}
\end{figure}

 Fig.~\ref{fig13} then summarizes our results for wall-attached crystal clusters, plotting contact angle $\theta$, basal radius $r$, particle
 number of the cluster $N^*$ and height $h$  versus the chosen total packing fraction $\eta$ in the system. Two different box sizes are used:
 as expected, $N^*$ (and also $h$ and $r$) must increase when the box size increases: as expected from Eq.~(\ref{eq13}), taking the box
 size to infinity at constant $\eta$ in the two-phase coexistence region we approach macroscopic phase coexistence. Approaching this limit,
 the contact angle $\theta$ should not change. Gratifyingly, we do find that the two data sets yield contact angles $\theta=70 \pm 2$ degrees
 irrespective of the choice of box size and packing fraction. Of course, for very small crystalline clusters (containing a few hundred particles only)
 a systematic effect on the contact angle is expected due to line tension effects, for the cluster sizes shown (where $N^*$ is several
 thousands) such effects are too small to be distinguished, given our statistical errors. Of course, in view of all the possible uncertainties
 whether or not $\gamma_{fc}$ depends on the interface orientation, and consequently whether a sphere cap shape is accurate, etc, the agreement
 of this finding for $\theta$ with the corresponding prediction based on the Young equation may be somewhat accidental. More work on this problem is
 desirable. 

\begin{figure} [ht]
\includegraphics[scale=0.313]{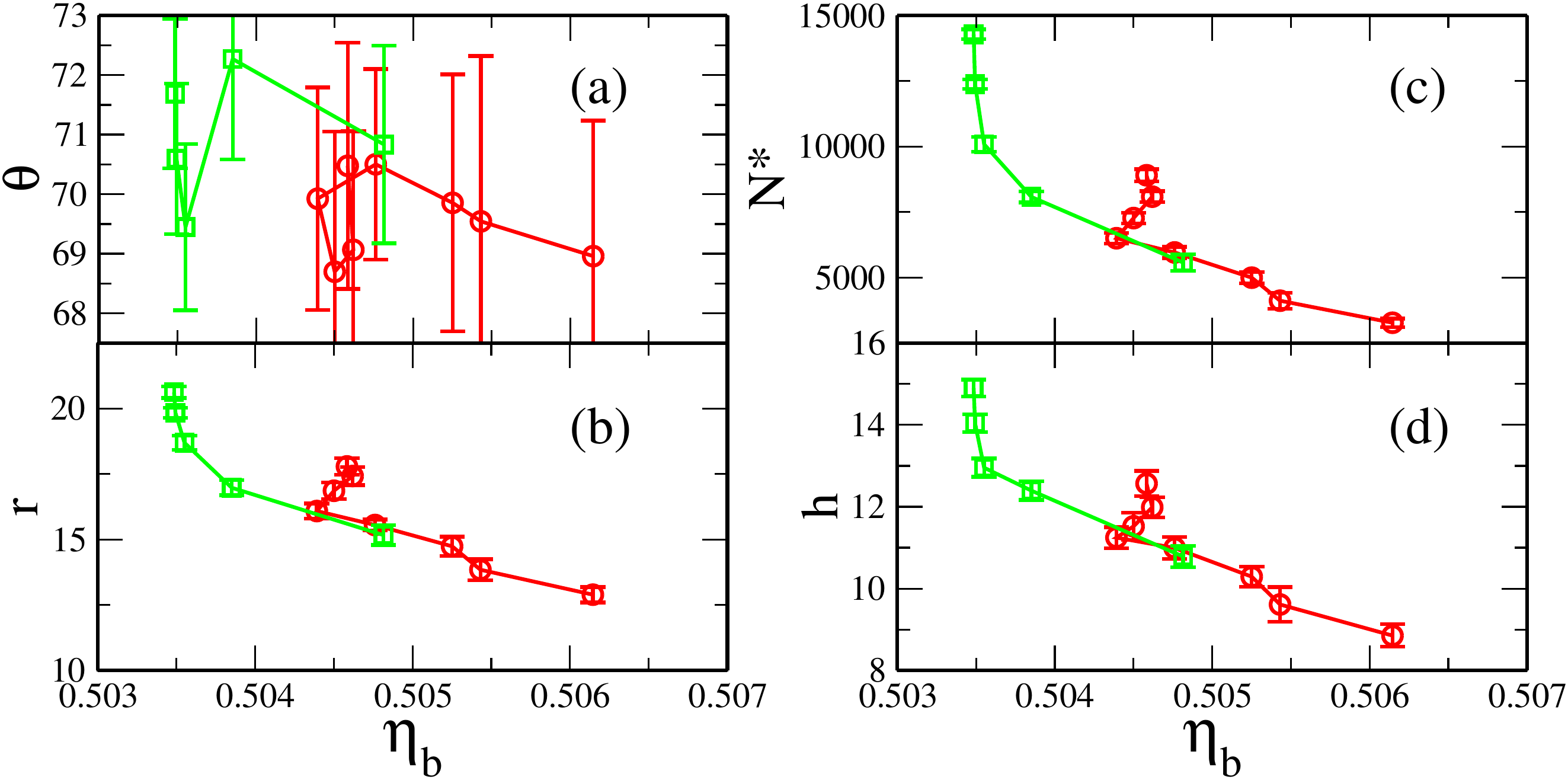}
\caption{\label{fig14} Same as Fig.~\ref{fig13}, but shown as functions of the packing fraction $\eta_b$ of the bulk metastable liquid coexisting with the crystalline clusters.}
\end{figure}

 In the studies of phase coexistence at the vapor-liquid transition \cite{14,15,45,46}, it was shown that it is also useful to analyze droplet
 properties as a function of the chemical potential of the vapor surrounding the liquid droplet, since then a part of the data collapses
 on ``master curves'' that are independent of the box linear dimensions (the parts that do not collapse are still affected by undesirable
 finite size effects due to the droplet evaporation-condensation transition for small droplets or by the transition from spherical
 to cylindrical shape for large droplets). In the present case of a liquid-solid transition, due to the high density of the liquid,
 the chemical potential of the liquid is not straightforwardly estimated, but we can easily determine the packing fraction $\eta_b$ of the bulk
 metastable liquid that coexists with the crystal cluster (this is done by sampling from a slab near $z=D/2$). Using the data of
 Fig.~\ref{fig13}, this is done in Fig.~\ref{fig14}. We expect $N^*$ (as well as $r$ and $h$) to decrease monotonically with increasing
 $\eta_b$ (the smaller the cluster, the denser the surrounding fluid must become, due to its pressure increase by the Laplace pressure which
 increases proportional to the inverse radius of curvature of the cluster). We see that most of the data do follow this expectation, but
 part of the data for the smaller system (for $N^*\geq 7500)$ break off from the common, size-independent ``master curve'' and also
 the corresponding contact angle data get systematically smaller. Similar trends were also seen when too large droplets were included into
 the analysis of the vapor to liquid transition \cite{14,15,45,46}: in this case it could be clearly proven that this trend is due to the problem
 that occasionally the droplet undergoes a transition in its shape from spherical to cylindrical \cite{45,46} (or sphere-cap to cylinder-cap
 in the presence of walls \cite{14,15}, respectively). This problem is easily missed without careful analysis of time evolutions of droplet
 properties, such as shown in Fig.~\ref{fig9},~\ref{fig10} (which refer to smaller $N^*$, where this effect did not yet occur). Similarly, also
 the data for the larger box size show a dramatic rise for $\eta_b \leq 0.5035$, where $N^*\geq 10 000$, and we feel that these data
 should be discarded for the same reason. Of course, this analysis of Fig.~\ref{fig14} is a first step only: it clearly would be desirable
 to extend this study to both smaller and larger box sizes, but in view of the huge demand in computer resources needed, this is
 left to future work. When we tentatively discard these data where $N^*$ is presumably too large, we obtain some evidence that $\theta$
 decreases with increasing $\eta_b$ (and hence decreasing $r$) slightly.

\begin{figure} [ht]
\includegraphics[scale=0.30]{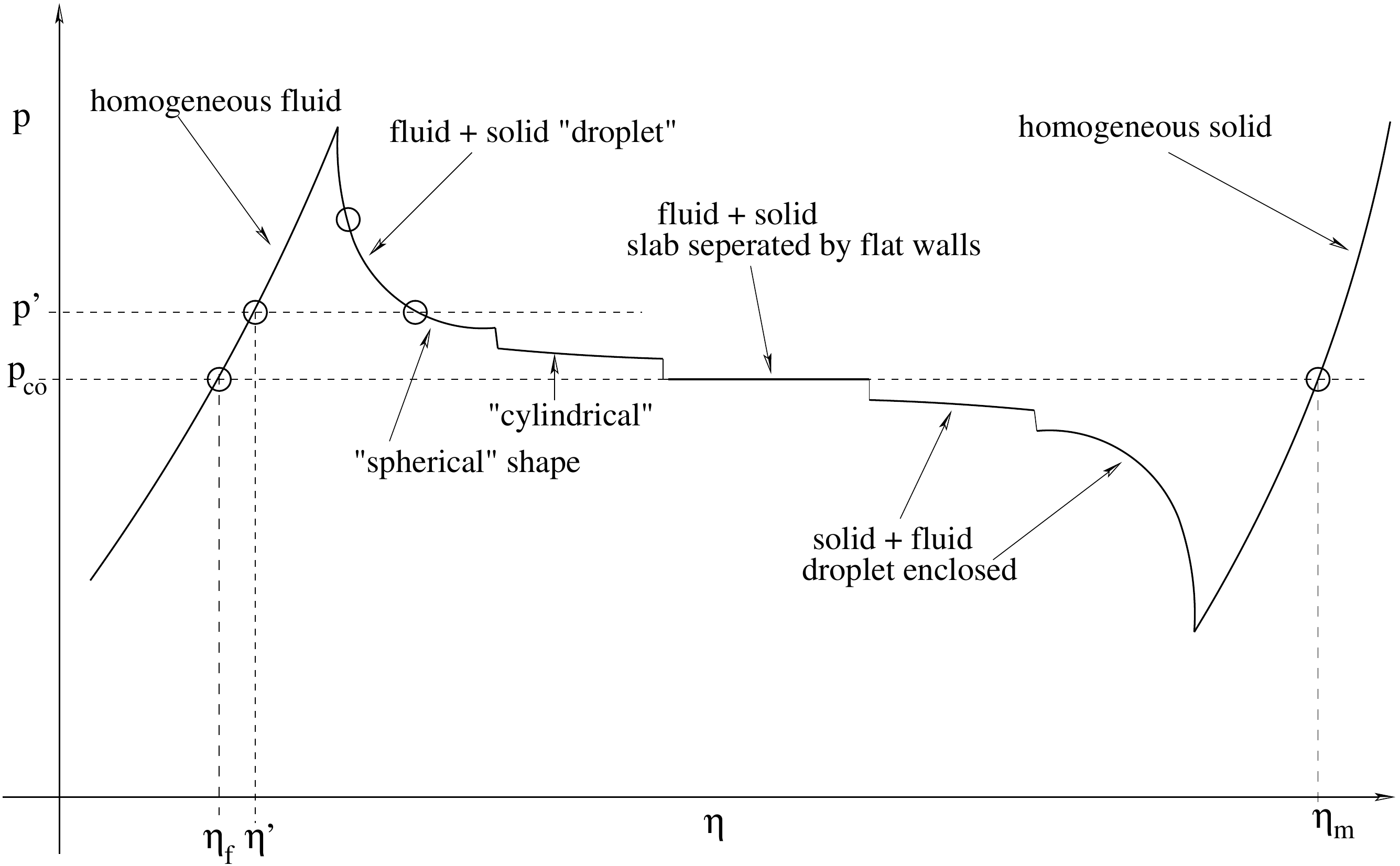}
\caption{\label{fig15}
Schematic description of the loop in the pressure versus packing fraction isotherm at a liquid-to-solid
transition in a finite simulation box. In the thermodynamic limit, $p$ rises with $\eta$ up to the value
$p_{co}$ at $\eta=\eta_f$, and stays constant at $p=p_{co}$ throughout the two-phase coexistence region,
until it continues to rise at $\eta=\eta_m$. In a finite system, the region of homogeneous fluid is enhanced
up to $\eta=\eta_t$, where the droplet evaporation/condensation transition occurs, and then it decreases smoothly, until another
transition occurs from spherical to cylindrical shape of the droplet. (In systems with walls, the actual droplet
shapes are sphere-cap like or cylinder-cap like, respectively). Only for the region of $\eta$ where the
two-phase coexistence in the finite box corresponds to a crystal slab separated by flat walls from the
fluid is the pressure (at least approximately) equal to $p_{co}$. For larger $\eta$ the roles of fluid and
crystal are interchanged. Note that in reality the transitions in Fig.~\ref{fig15} are not sharp but slightly rounded. 
When the volume of the system tends to infinity, $\eta_t\rightarrow \eta_m$, and the pressure enhancement 
$p(\eta_t)-p_{co}\rightarrow0$, so while the extrema of the loop in Fig.~\ref{fig15} gets the sharper the 
larger the system becomes, the whole loop ultimately has disappeared after the thermodynamic limit has been taken.}
\end{figure}

 From the knowledge of the contact angle $\theta$ and of the radius $R^* $ of a droplet one can make a prediction of the
 corresponding free energy barrier $\Delta F^*_{\rm het}$ that needs to be crossed, if such a droplet forms
 spontaneously by thermal fluctuations.

 However, experience with nucleation in vapor-liquid transitions or fluid-fluid unmixing \cite{14,15,45,46,47}
 suggests that barriers predicted from Eqs.~(\ref{eq11}),~(\ref{eq12}) often are significantly larger
 than the actual barriers. In order to test whether this problem also occurs in the present case,
 we recall that classical nucleation theory predicts also a relation for the pressure $p'$ of the liquid
 coexisting with the droplet, namely \cite{1,2} (cf. Fig.~\ref{fig15} for notation; Fig.~\ref{fig15} is a
 schematic counterpart of Fig.~\ref{fig3})

 \begin{equation} \label{eq17}
 p'-p_{co}= (2 \gamma_{fc} ) / R^* \quad.
 \end{equation}

 Using our estimate for $R^*$ and $\gamma_{fc}$ from \cite{40} one predicts that for $R^*=14.4$
 the pressure difference is $p'-p_{co} \approx0.13$ while the actual pressure difference seen for this case in Fig.~\ref{fig3} is
 $p'-p_{co} \approx0.4$. Potential causes for this failure include the assumed spherical shape of the crystalline cluster in eq.~\ref{eq17} and the exact definition of $R^*$ in our simulation. Clearly, more work is required to study the surface free energy of the curved crystalline clusters.

\section{Conclusions}
In this paper we have presented a study of liquid-solid phase coexistence in the constant volume (NVT) ensemble for the Asakura-Oosawa model of a colloid-polymer mixtures, focusing on the case of the size ratio $q=0.15$ and polymer reservoir packing fraction $\eta^r_p=0.1$, for which the bulk phases and the interfacial stiffness $\gamma_{100}$ for a fluid-solid interface with an (100) surface of the crystal were studied in previous work \cite{40}. Using rather large systems (containing of the order of almost 10$^5$ colloidal particles), we have shown that an analysis based on lever-rule type arguments allows us to gain insight on many aspects of phase coexistence, both with respect to bulk properties characterizing it, and with respect to the contact angle of sphere-cap shaped crystalline clusters. When we choose the packing fraction $\eta$ roughly half way in between the bulk coexisting liquid $(\eta_f)$ and solid $(n_m)$ phases, we obtain a phase coexistence, where adjacent to the left wall there is a crystal film (with (111) planes stacked on top of each other up to a distance $z=D_l$), followed by a liquid up to the distance $D-D_r$, and then another crystal film up to the right wall (at distance $D$ from the left wall) follows.
We have checked (Appendix A) that this liquid slab that forms in between these crystal layers is in full thermal equilibrium and its packing fraction $\eta_f$ and pressure $p_{co}$ can be measured very precisely in this way. One can show that these values are independent of $\eta$ (and hence direct evidence for the flat horizontal portion of the pressure vs. -packing fraction isotherm, Fig.~\ref{fig3}, is provided), and from the observation of the profile $\eta(z)$, Fig.~\ref{figA1}, one can also measure $D_l + D_r$ and thus establish that the lever rule holds, as it should be. From the profiles one can also extract the distance $d$ between the crystalline (111) planes and thus check the self-consistency of the estimation of $\eta_m$.

The main interest of this paper, was the study of wall-attached crystal clusters (Figs.~\ref{fig5},~\ref{fig6},~\ref{fig9}-\ref{fig11},
\ref{fig13},~\ref{fig14}), which were created by special choice of initial states (Fig.~\ref{fig4}) for packing
fractions $\eta$ that exceed $\eta_f$ only slightly. If this excess is too small, such crystal clusters were found to be unstable
and dissolve again, and one is left with a somewhat compressed uniform fluid (apart from the layering at the walls, similar to
what is seen in Figs.~\ref{figB1},~\ref{figB2} at the right wall). If this excess is too large, crystalline clusters of cylindrical shape
(or even planar crystalline films) form (Fig.~\ref{fig7}). Using the liquid-wall and crystal-wall surface tensions that were
estimated by a different method \cite{64} and are shown in Fig.~\ref{fig12}, one can estimate the contact angle to be close
to 70$^o$ (this estimate relies on the assumption that the fluid-crystal interface tension $\gamma_{fc}$ is approximately
independent of crystal surface orientation, and hence the estimate of \cite{40} can be used here). Gratifyingly, the numerical
data observed for the contact angle $\theta$ (Figs.~\ref{fig9},~\ref{fig10},~\ref{fig13},~\ref{fig14}) are compatible with this
estimate. Of course, the crystal-wall interfacial tension $\gamma_{wc}$ is expected to depend significantly on the
orientation of the crystal axes relative to the wall. Consistent with this expectation, a contact angle close to
90$^o$ is observed (Fig.~\ref{figB1}) if (100) planes of the crystal are chosen to be parallel to the wall surface. The estimate of 
$\gamma_{wc}$ for the (100) orientation from the "ensemble mixing" method is $\gamma_{wc}^{100} = 1.82 \pm 0.051$ which leads to a contact
angle of $\theta \approx 90^o$.

We have also pointed out that in the finite system the pressure versus packing fraction isotherm (Figs.~\ref{fig3},~\ref{fig13})
exhibits a loop, which has nothing whatsoever to do with van der Waals-like loops; however: it is entirely caused by
interfacial contributions to the free energy of the system, and since the latter are down by a surface to volume-ratio
in comparison with the bulk, the loop gradually develops towards a flat variation $p=p_{co}$ from $\eta=\eta_f$ to
$\eta=\eta_m$, when the thermodynamic limit is taken. Since the actual pressure enhancement (in the region where
$p$ decreases with increasing $\eta$, Figs.~\ref{fig3},~\ref{fig13}) should contain information on the interfacial
free energies of the (spherical or cylindrical crystalline clusters), we have tried to study this enhancement of the pressure, too,
but using Eq.~(\ref{eq17}) did not yield results that were quantitatively consistent with our other results. This
problem clearly requires further study.

We like to emphasize, that the present work is a first step only; it is necessary to develop thermodynamic integration-based
methods, from which the excess free energy of the system (due to the crystalline cluster) can be reliably extracted, in order to be able
to make quantitative predictions for nucleation barriers in such systems. Also it will be interesting to study the same
model for other values of $q$ and $\eta^r_p$. In any case, it would also be very interesting if experiments on colloid-polymer
mixtures, which are rather well described by this model, were performed.

\underline{Acknowledgement}. We acknowledge support by the Deutsche Forschungsgemeinschaft (DFG) under grants No Bi 314/19-2
and SFB TR 6/A5, and thank the John von Neumann Institute for Computing (NIC) for a grant of computer time. We are grateful
to J. Horbach, M. Oettel and A. Troester for useful discussions, and thank in particular T. Zykova-Timan for information
on the codes used for the computations published in Ref. \cite{40}.

\appendix

\section{Phase coexistence between wall-attached crystal films with a liquid slab in between}

When we prepare a system at a packing fraction about halfway in between the packing fractions $\eta_f$, $\eta_m$
of the bulk coexisting phases, we expect that the final equilibrium state will be a layered structure, with
a crystal film of thickness $D_l$ attached to the left wall, a crystal film of thickness $D_r$ attached to the
right wall, and a fluid slab of thickness $D-D_r-D_l$ in between. From symmetry, one expects $D_r=D_l$,
of course, and the total thickness of the crystal films $D_r + D_l$ is fixed by the lever rule,

\begin{equation} \label{eqA1}
D\eta = (D-D_r-D_l) \eta_f + (D_r + D_l) \eta_m \quad.
\end{equation}

\begin{figure} [ht]
\includegraphics[scale=0.700]{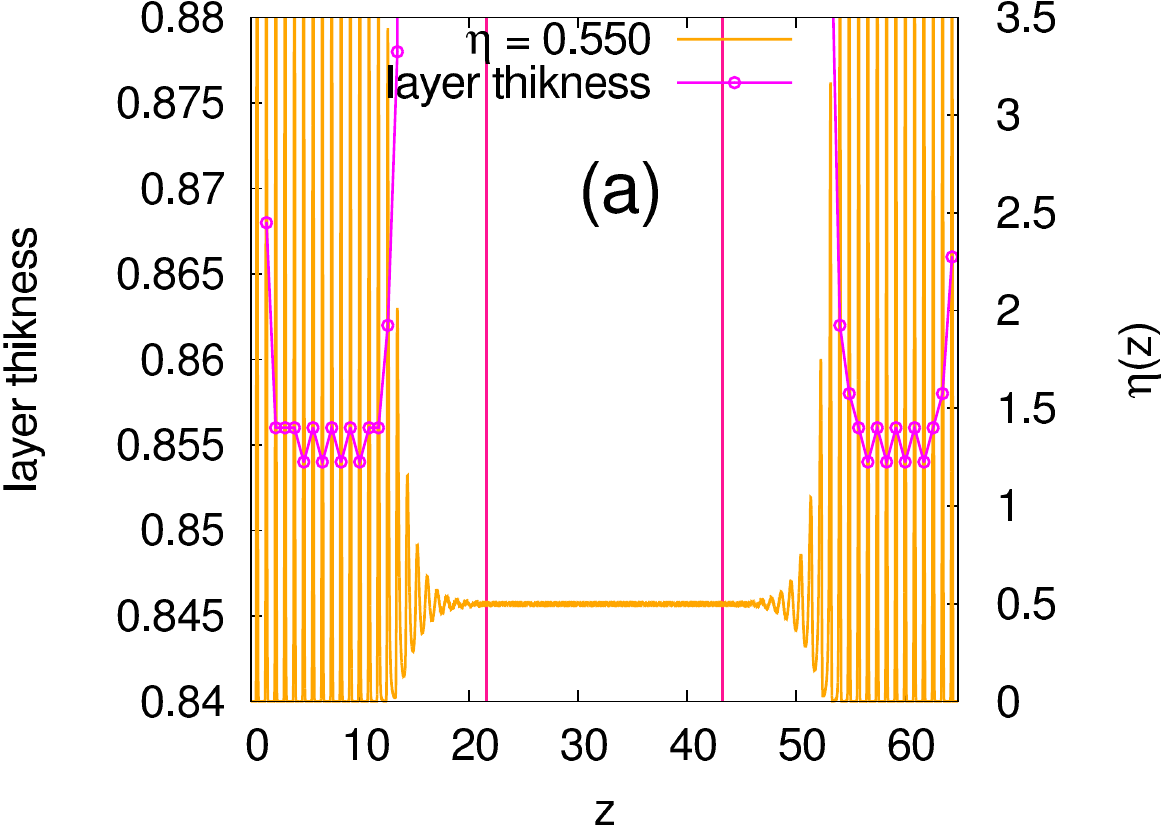}
\includegraphics[scale=0.700]{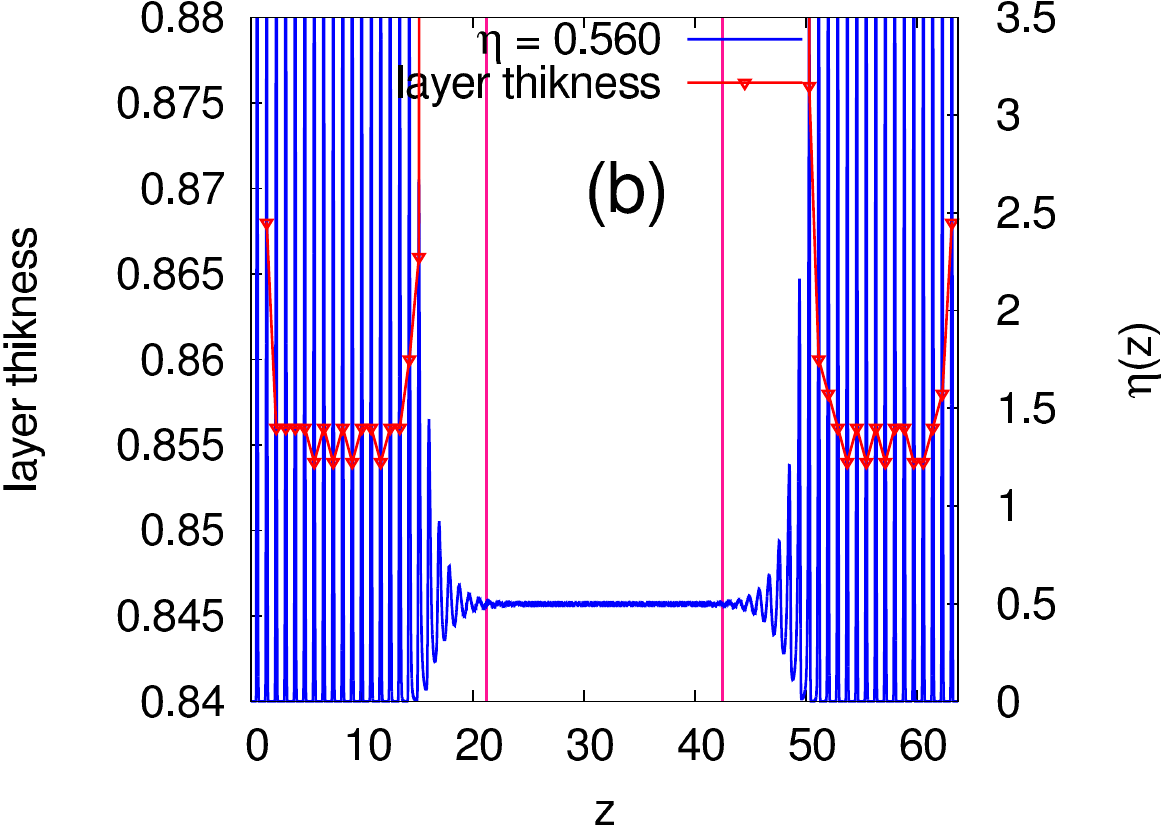}
\caption{\label{figA1} Plot of the packing fraction profile $\eta (z)$, ordinate labels on the right side of the picture, and of
the local estimates $d(z)$ for the distance between neighboring planes in the fcc ABCAB$\cdots$ stacking,
ordinate labels on the left side of the figure, for the choice of linear dimensions $L_x=33.59396$, $L_y=29.09322$,
$D=64.83302$, $\eta=0.55$(a) and $D=63.67528$, $\eta=0.56$(b). The thick vertical lines in the center of these
figures mark the part of the liquid slab that has been used for the estimation of $\eta_f$ and $p_{co}$. Note that
this figure shows only two examples out of many more that were recorded.}
\end{figure}

Such a state is easily obtained if we first prepare the total $L_x \times L _y \times D$ system as a
crystal of packing fraction $\eta_m$ (choosing linear dimensions $L_x$, $L_y$, $D$ such that any misfit
with the fcc crystal structure at $\eta_m$ is strictly avoided), and removing then particles from the region
near $z=D/2$ from the system, until the desired average packing fraction is obtained. Then the system is
equilibrated, and one observes that quickly a liquid slab forms in between the crystal layers
(Fig.~\ref{figA1}). The thicknesses of the two crystal layers $D_r$, $D_l$ are only roughly equal
to each other, but as long as each crystal contains many (111) layers stacked parallel to the wall, 
no noticeable systematic error is caused by this slight asymmetry. One sees that
irrespective of the precise value of $\eta$ that is chosen, one does obtain an extended region
near $z=D/2$ where the volume fraction profile $\eta(z)$ is flat, and this horizontal region yields
an accurate estimate for $\eta_f$. Note that for a wide range of choices of $\eta (0.53 \leq \eta \leq 0.58)$
and also for several choices of the linear dimensions we always obtain the same value of $\eta_f \approx 0.498$
(slightly larger than the estimate $\eta_f= 0.494$ of Zykova-Timan et al. \cite{40}). However, the coexistence
pressure $p_{co}$ (extracted from the region when $\eta (z)$ in Fig.~\ref{figA1} is flat, using the method
described in \cite{70}) agrees with the previous estimation \cite{40} $p_{co}=8.00 \pm 0.01$ within the statistical
errors (the present estimate is $p_{co}=7.98 \pm 0.01$). Starting from the initial estimate for the
packing fraction of the crystal $\eta_m=0.64$, one obtains for the distance between the crystal (111) planes $d=0.857$.
Within the accuracy with which $d$ can be estimated from the profiles $\eta(z)$ in the crystalline films, this is the value
that the simulation yields, Fig.~\ref{fig11}. Of course, one cannot reliably measure $d$ right at the walls, and one
should also avoid using the profiles in the region of the crystal-liquid interface. Estimating $D_l + D_r$ from the
location of the interface positions in the profiles $\eta(z)$, we have confirmed Eq.~(\ref{eq11}) quantitatively.

This type of crystal-liquid coexistence simulation hence not only provides direct evidence for the strictly horizontal
part of the pressure versus packing-fraction isotherm in Fig.~\ref{fig15}, but yields direct estimates for
$\eta_f$, $\eta_m$ and $p_{co}$ with very good precision. Of course, we have simplified matters by using the
(previously known) value of $\eta_m$ to choose linear dimensions $L_x$, $L_y$ commensurate with the fcc lattice
structure that the system wants to develop. If the initial choice of $\eta_m$ would be somewhat off, one
would find that the crystal structure exhibits some elastic distortion: the distance between planes would come
out either somewhat larger or smaller than predicted from the (wrongly chosen) initial value for $\eta_m$:
then an iterative improvement of this choice would be necessary. Thus, we propose such studies of phase
coexistence as an additional method to precisely characterize liquid-solid transitions.

\section{Simulation of metastable crystals with (100) planes oriented parallel to the walls}

\begin{figure} [ht]
\includegraphics[scale=0.600]{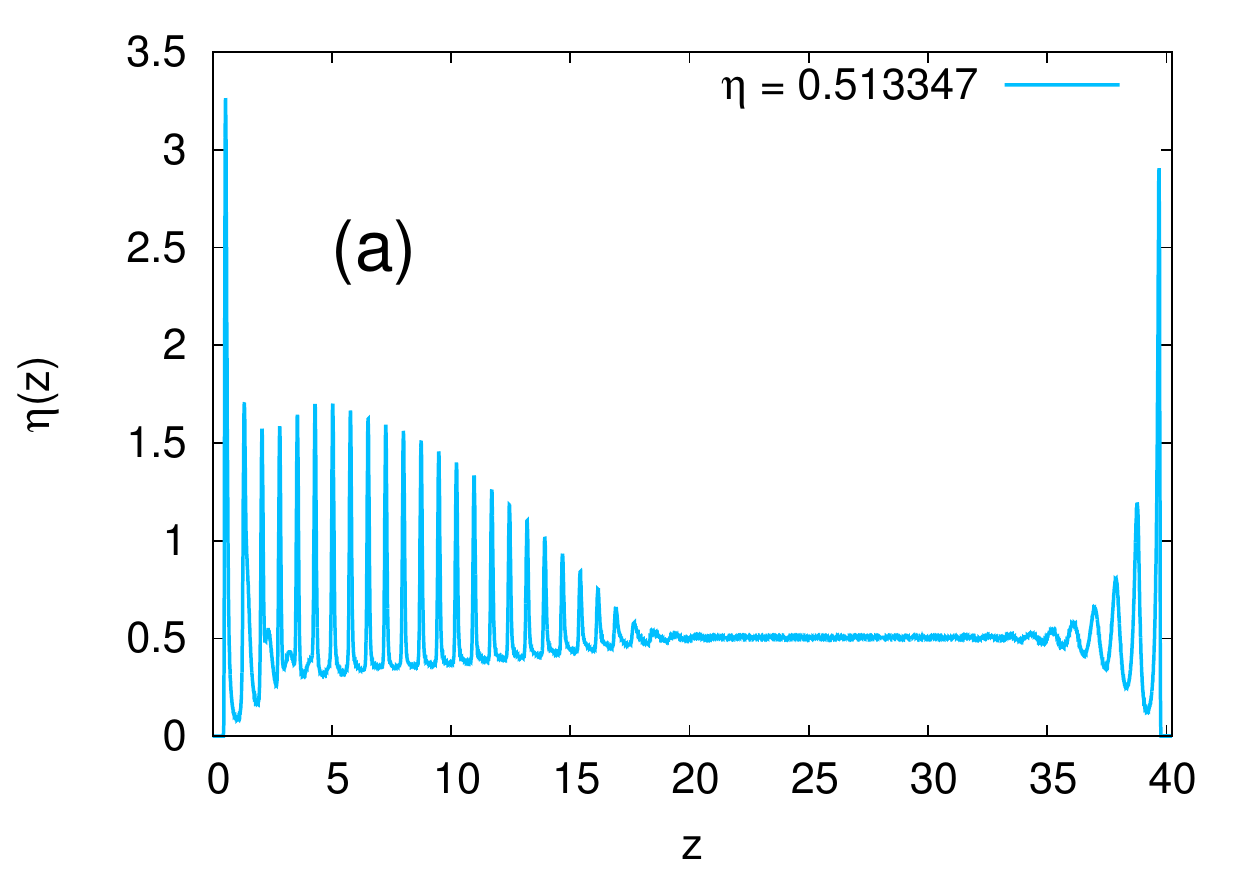}
\includegraphics[scale=0.70]{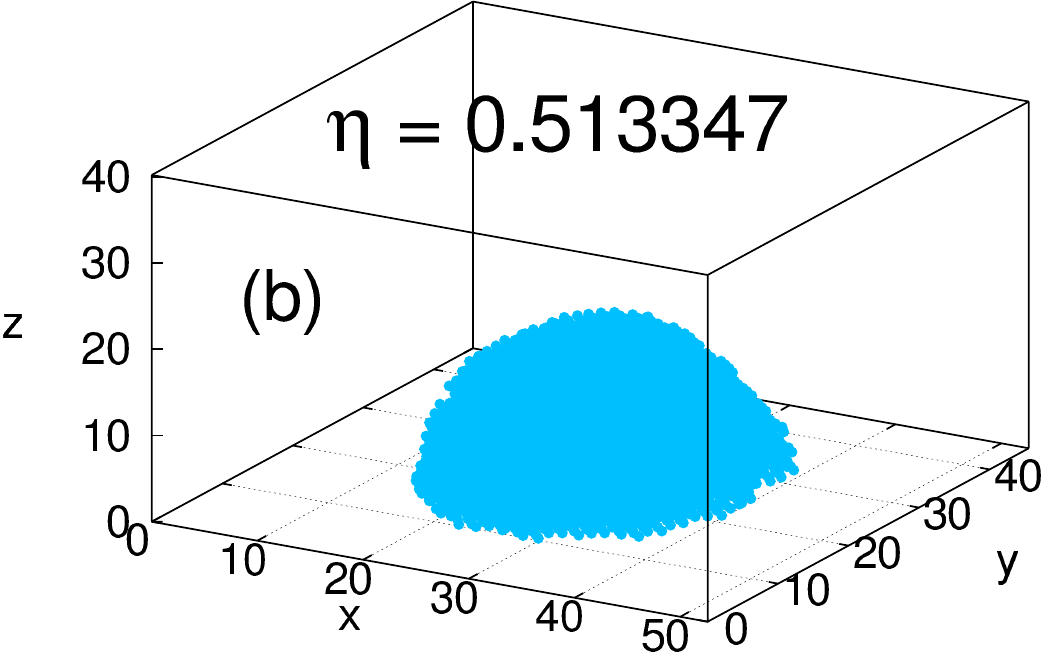}
\includegraphics[scale=0.15]{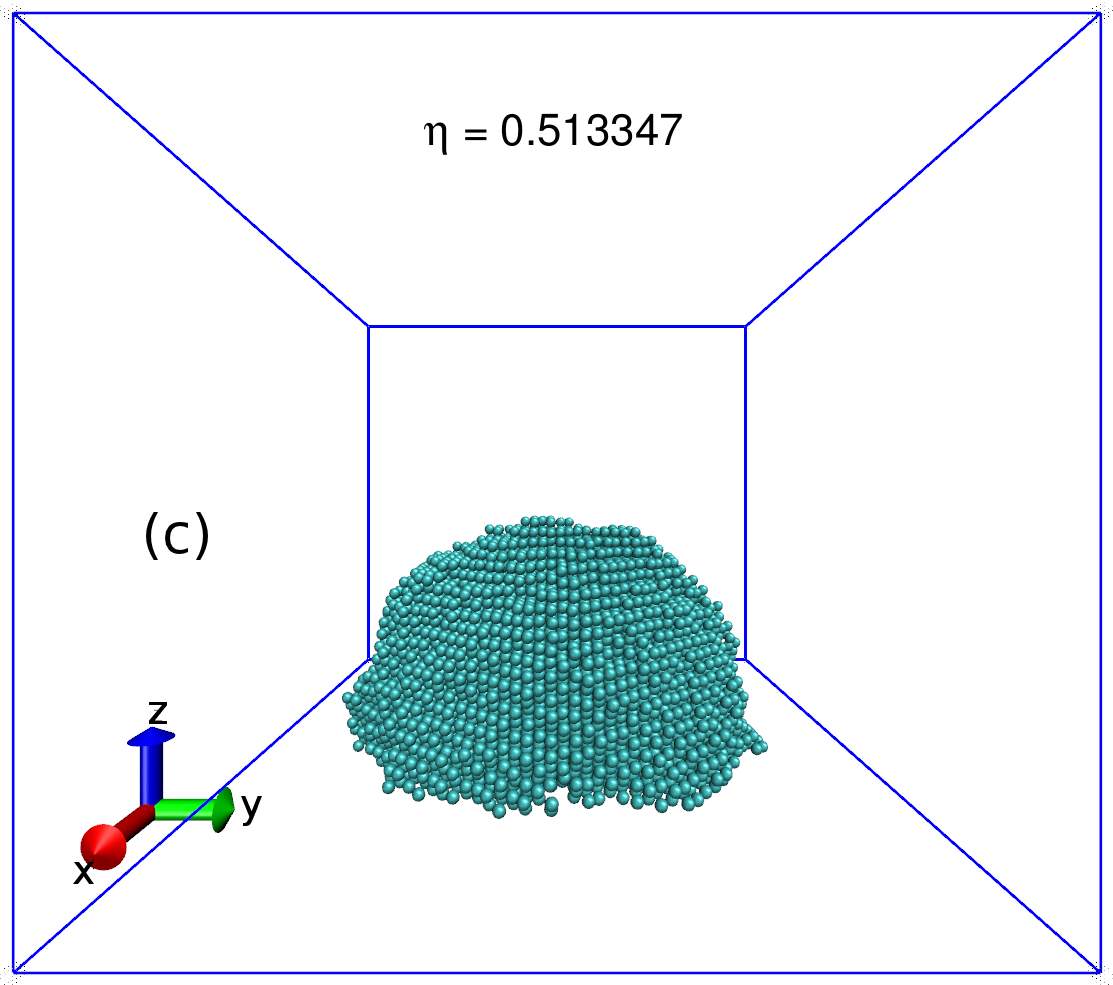}
\caption{\label{figB1} (a) Packing fraction profile and (b)-(c) two different perspectives  of a crystalline cluster with (100) plane stacking for $\eta=0.513347$}.
\end{figure}

\begin{figure} [ht]
\includegraphics[scale=0.600]{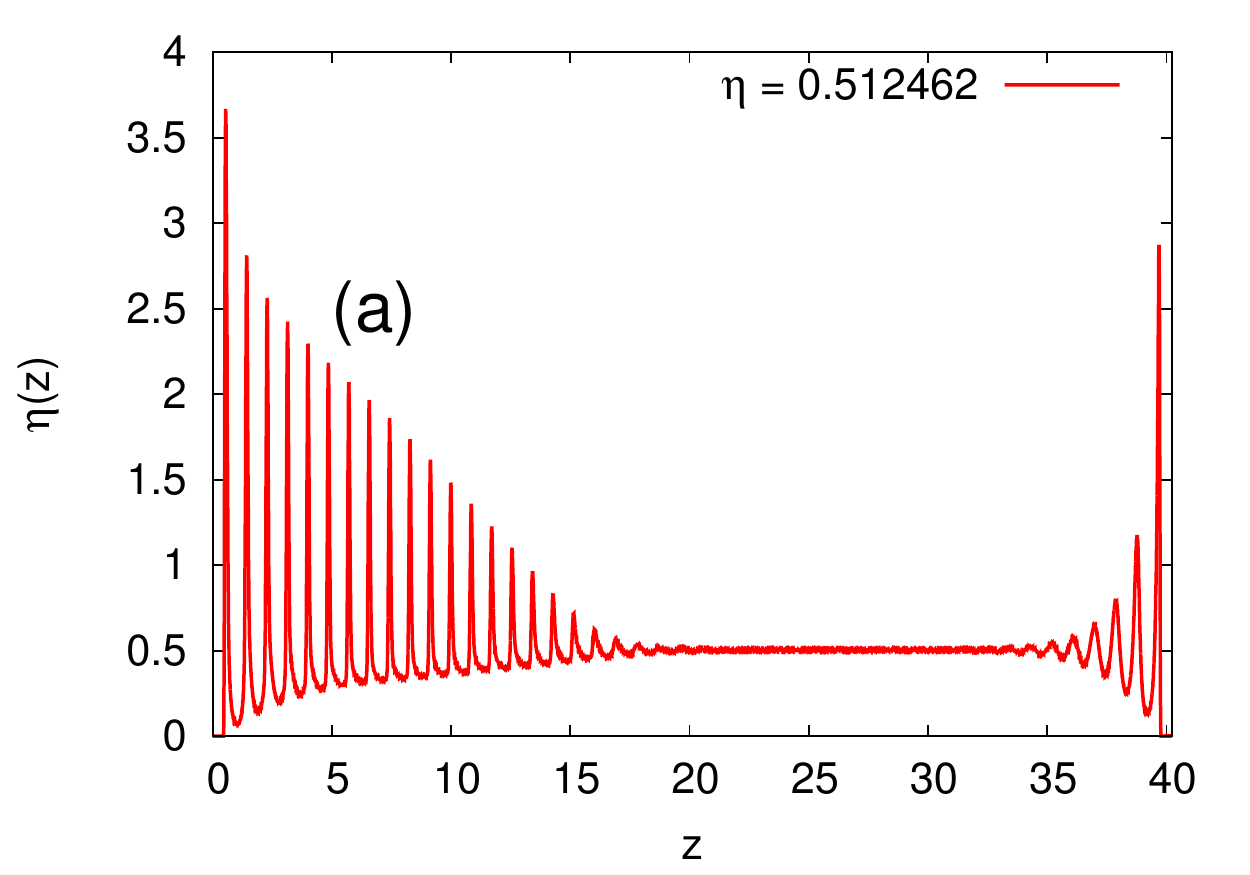}
\includegraphics[scale=0.70]{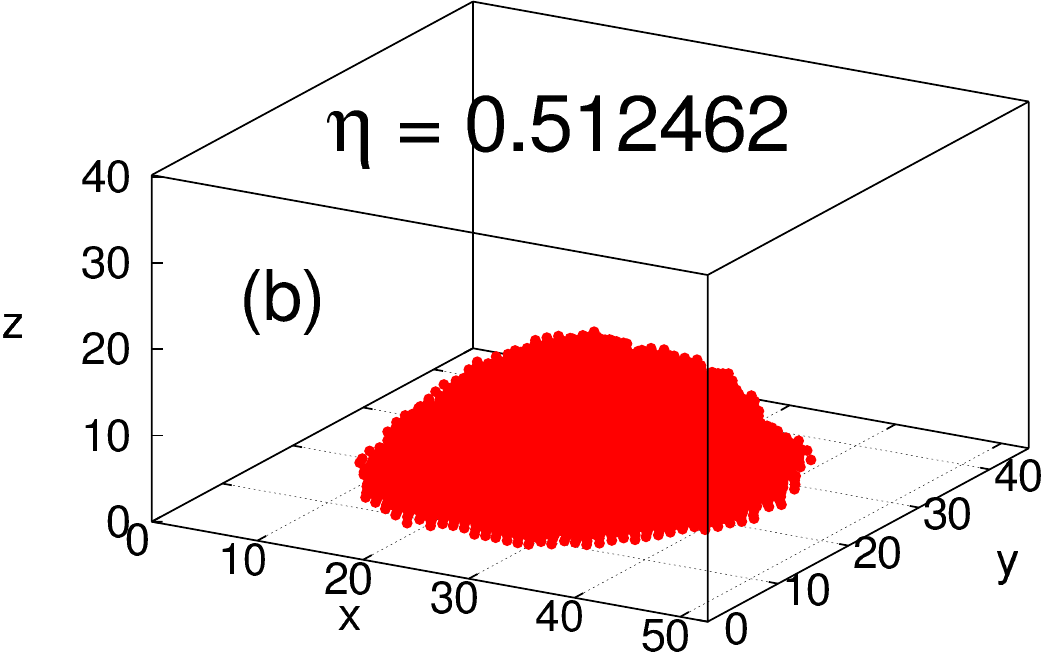}
\includegraphics[scale=0.15]{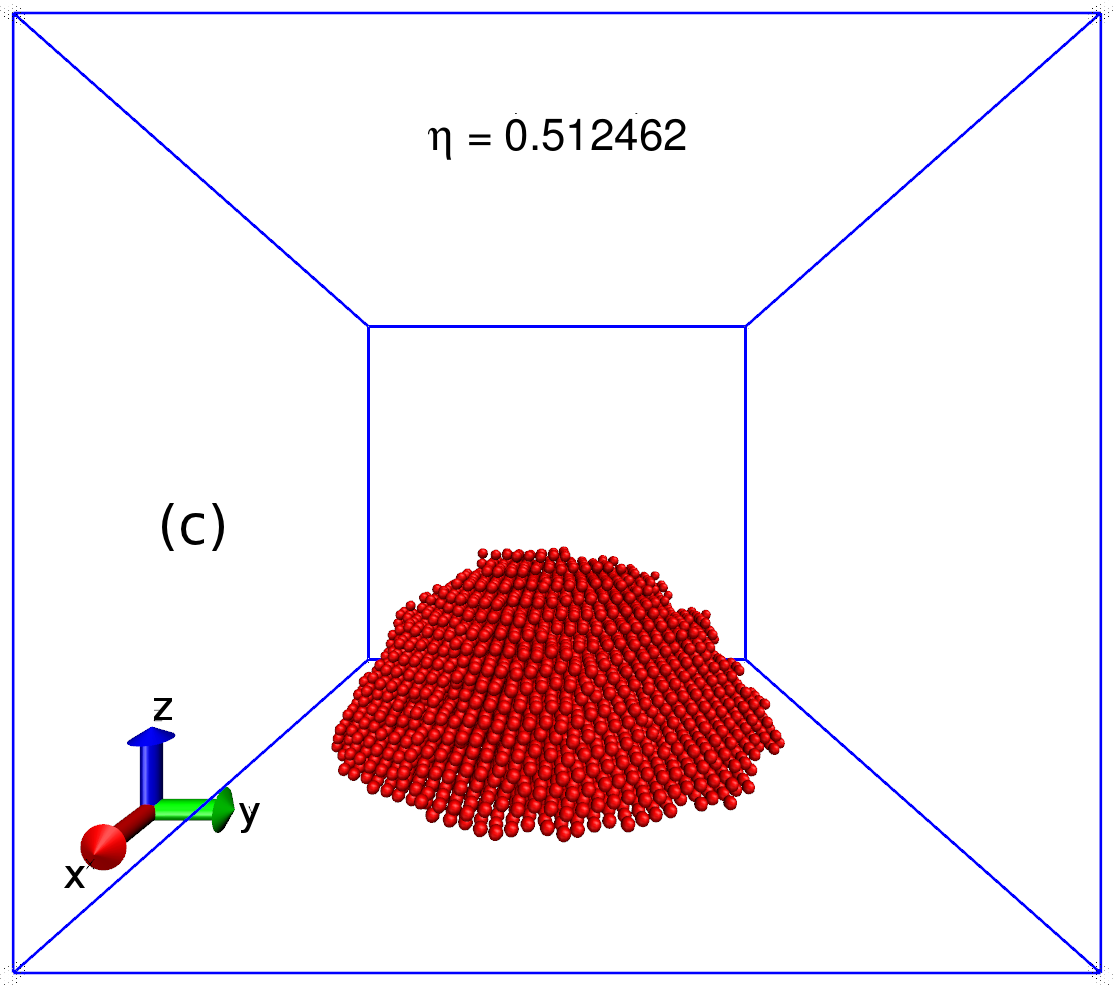}
\caption{\label{figB2} Same as Fig.~\ref{figB1}, but for (111) stacking and $\eta=0.512462$}.
\end{figure}

In the main text, we have described the technique to prepare crystalline clusters where by construction of the initial states
the familiar ABCABC stacking of the close-packed (111) planes parallel to the planar walls ware created. It is possible, however,
to choose initial states where instead the (100) planes are stacked parallel to the planar walls. While in the (111) planes the particles form a triangular
lattice, where (for $\eta_m=0.64$) the lattice spacing is 1.0497, and the distance between planes is $d=0.857$, for the
(100) planes the particles form a square lattice, with the same lattice spacing, but the distance between the planes is
slightly smaller, than for the (111) stacking, namely $d=0.7423$. Figs.~\ref{figB1},~\ref{figB2} compare typical cases  of
clusters with (100) stacking and (111) stacking. Note that at the right wall ($z$ close to $D$) there is the typical
layering of the liquid phase near a flat repulsive wall, which is very similar in both cases. In the profile near the
left wall, the first peak of $\eta(z)$ adjacent to the wall is again in part due to the layering of the fluid and in part due to
the crystalline cluster, and again similar in both cases. However, while for (111) stacking the further density oscillations (which are
mostly due to the crystalline cluster) decrease monotonically with the distance $z$ from the wall, this is not the case for (100)
stacking: the $3^{\rm rd}$ and 4$^{\rm th}$ peak of the oscillations are less high than the 5$^{\rm th}$ to 8$^{\rm th}$ peaks.
This non-monotonic behavior of the peak heights has an obvious interpretation in terms of the cluster shapes: for (100)
stacking the contact angle exceeds 90$^o$ slightly, and so the cross-sectional area of the crystalline cluster along the 5$^{\rm th}$ to
8$^{\rm th}$ plane is slightly larger than along the 3$^{\rm rd}$ and 4$^{\rm th}$ plane. The bottom snapshot gives direct
visual evidence for this interpretation. Due to the larger contact angle this crystalline cluster is only metastable.

\end{document}